\documentclass[
aip,
amsmath,amssymb,
preprint
]{revtex4-2}

\usepackage{natbib}
\usepackage{graphicx}
\usepackage{mathptmx}
\usepackage{hyperref}
\usepackage{nameref}
\usepackage[utf8]{inputenc}
\usepackage[T1]{fontenc}
\usepackage{etoolbox}

\begin{document}

\title{Fragmentation of Water Clusters Formed in Helium Nanodroplets by Charge Transfer and Penning Ionization}

\author{S. De}
\affiliation{Quantum Center of Excellence for Diamond and Emergent Materials and Department of Physics, Indian Institute of Technology Madras, Chennai 600036, India}

\author{A. R. Abid}
\affiliation{Department of Physics and Astronomy, Aarhus University, 8000 Aarhus C, Denmark}

\author{J. D. Asmussen}
\affiliation{Department of Physics and Astronomy, Aarhus University, 8000 Aarhus C, Denmark}

\author{L. Ben Ltaief}
\affiliation{Department of Physics and Astronomy, Aarhus University, 8000 Aarhus C, Denmark}

\author{K. Sishodia}
\affiliation{Quantum Center of Excellence for Diamond and Emergent Materials and Department of Physics, Indian Institute of Technology Madras, Chennai 600036, India}

\author{A. Ulmer}
\affiliation{Department of Physics, Universit{\"a}t Hamburg, Luruper Chaussee 149, 22761 Hamburg, Germany}

\author{H. B. Pedersen}
\affiliation{Department of Physics and Astronomy, Aarhus University, 8000 Aarhus C, Denmark}

\author{S. R. Krishnan}
\email{srkrishnan@iitm.ac.in}
\affiliation{Quantum Center of Excellence for Diamond and Emergent Materials and Department of Physics, Indian Institute of Technology Madras, Chennai 600036, India}

\author{M. Mudrich}
\email{mudrich@phys.au.dk}
\affiliation{Department of Physics and Astronomy, Aarhus University, 8000 Aarhus C, Denmark}

\date{\today}

\begin{abstract}
Helium nanodroplets (``HNDs'') are widely used for forming tailor-made clusters and molecular complexes in a cold, transparent, and weakly-interacting matrix.
Characterization of embedded species by mass spectrometry is often complicated by fragmentation and trapping of ions in the HNDs.
Here, we systematically study fragment ion mass spectra of HND-aggregated water and oxygen clusters following their ionization by charge transfer ionization (``CTI'') and Penning ionization (``PEI'').
While the efficiency of PEI of embedded clusters is lower than for CTI by about factor 10, both the mean sizes of detected water clusters and the relative yields of unprotonated cluster ions are significantly larger, making PEI a ``soft ionization'' scheme.
However, the tendency of ions to remain bound to HNDs leads to a reduced detection efficiency for large HNDs containing $>10^4$ helium atoms.
These results are instrumental for determining optimal conditions for mass spectrometry and photoionization spectroscopy of molecular complexes and clusters aggregated in HNDs.
\end{abstract}

\maketitle 

\section{Introduction}
Water (H$_2$O) is a crucial component for all living organisms ~\cite{lo_water_2000,laurson_water_2015}
and the Earth's atmosphere.~\cite{falcinelli_possible_2015,vaida_perspective_2011,douroudgari_impact_2022}
Therefore, studying the fundamental properties of water, both in the liquid phase and in the vapor phase, has been a major line of research in natural sciences over many decades. 

Clusters are an intermediate state of matter bridging the gap between individual molecules and the condensed phase.
Clusters allow us to study the properties of condensed-phase systems using sophisticated gas-phase diagnostic techniques such as ion mass spectrometry and ion-momentum and electron spectroscopy.
The understanding of water clusters and their interaction with ionizing radiation has fundamental implications for radiation damage in biological systems~\cite{kohanoff_interactions_2017,ren_experimental_2018} and for the photochemistry of the atmosphere.~\cite{anglada_atmospheric_2013}

Ionization of water clusters in the gas phase leads to the formation of protonated water-cluster ions by intracluster proton transfer and dissociation.
Significant structural changes occur due to excess energy deposited in the cluster ions after their vertical ionization, which leads to the rearrangement of the cluster and subsequent loss of a hydroxyl (OH) radical ~\cite{tachikawa_ionization_2004}
   \begin{eqnarray}
(\mathrm{H_2O})_k~+~h\nu(\mathrm{or}~e^-)~\rightarrow~ (\mathrm{H_2O})_k^+~+~e^- (\mathrm{or}~2e^-)\\
(\mathrm{H_2O})_k^+~\rightarrow~\mathrm{H}^+(\mathrm{H_2O})_{k-m-1}~+~m\cdot \mathrm{H_2O}+\mathrm{OH}
   \end{eqnarray}
         
Usually, protonated water cluster ions are the most abundant products after the ionization of water clusters compared to ``unprotonated'' water cluster ions, where the proton transfer (``PT'') occurs, but OH radical is still attached to the hydronium ion.
The PT time in water dimer cation has been directly measured by Schnorr~\emph{et al.} recently using XUV pump and XUV probe measurements, yielding a PT time of $(55 \pm 20 )$ fs.~\cite{schnorr2023direct} Therefore, the formation of ``unprotonated'' water cluster ions requires a process that competes with PT by rapidly removing the excess energy deposited in the cluster by vertical ionization.
This ultrafast quenching is only possible by the presence of a second atom or molecule. Shinohara~\emph{et al.} observed ``unprotonated'' water cluster ions in ion mass spectra for the first time.
Fragmentation was quenched by attaching Ar atoms which helped to remove the excess energy by evaporation of some Ar atoms.
Thus, evaporation of Ar atoms constitutes a rapid cooling process.~\cite{shinohara_photoionization_1986}
Jongma \emph{et al.} experimentally and theoretically determined the structure of ``unprotonated'' water cluster ions H$_3$O$^+$(H$_2$O)$_{k}\cdot$OH where the OH radical is located outside of the first solvation shell.~\cite{jongma_rapid_1998}
Iguchi \emph{et al.}\cite{iguchi2023isolation} very recently identified the hemibonded structure of water dimers, where two H$_2$O share the unpaired electron and the excess charge.
Thus, in ``unprotonated'' water cluster ions, there are two competing forms, one of which is proton transferred H$_3$O$^+\cdot$OH and another is hemibonded (H$_2$O$\cdot$OH$_2$)$^+$.
Although the hemibonded structure is unstable due to its comparatively high energy, it can be observed when rapidly cooling the complex, \textit{e.~g.} by embedding it in a cold environment. 

Ultrafast quenching of PT and fragmentation occurs in embedded systems in HNDs, which are considered nearly ideal host matrices.
Due to their ultra-low equilibrium temperature ($0.4$~K), embedded water clusters are rapidly cooled, and the formation of protonated water cluster ions is suppressed.
Yang~\emph{et al.}~\cite{yang_electron_2007} reported electron impact ionization mass spectra of water clusters in HNDs, and they observed He(H$_2$O)$_k^+$ ions as well as pure protonated and "unprotonated" water cluster ions.
They attributed the formation of He(H$_2$O)$_k^+$ to the structure and charge distribution of molecular water cluster cations; He(H$_2$O)$_k^+$ clusters are produced as a result of the formation of (H$_2$O)$_k^+$ ions with an exposed H$_3$O$^+$ unit at the cluster surface with a single dangling O-H bond to which a single helium atom can attach via charge-induced dipole interaction.
A remarkable softening effect of the ionization was observed when a co-dopant is used along with water clusters.~\cite{liu_coreshell_2011, denifl_ionization_2010, denifl_ionmolecule_2009} Co-dopants (N$_2$, O$_2$, CO$_2$, and C$_6$H$_6$) create a protective shell surrounding the core of the water cluster, effectively inhibiting the formation of protonated cluster ions.
However, the presence of co-dopants (CO and NO) can also promote the fragmentation of specific water cluster sizes through intricate secondary reactions.~\cite{liu_coreshell_2011}

Superfluid HNDs are extensively used as cryogenic nanometer-sized quantum host matrices due to their intriguing properties such as ultralow temperature, vanishing viscosity, and chemically inert environment.
They can easily pick up any foreign atoms and molecules, which are rapidly cooled to $\sim 0.4$~K.
Dopant species can then aggregate and form atomic or molecular clusters and even nanostructures.~\cite{boatwright_helium_2013, haberfehlner_formation_2015}
When HNDs are irradiated by electrons or extreme ultraviolet radiation (``EUV''), HNDs no longer act as inert cryo-matrices, but themselves become highly reactive species.
In the regime of resonant excitation at photon energies $20.5 < h\nu < 23$~eV, HNDs are excited into an excited state correlating to the $1$s$2$s and $1$s$2$p configurations of atomic helium, which has a large absorption cross-section of about 25~Mbarn per helium atom.~\cite{buchta2013extreme}
The excitation tends to relax by localizing on a single excited atom which is expelled toward the surface on a ps time scale.~\cite{Ziemkiewicz_ultrafast_2015,mudrich_ultrafast_2020}
In case a foreign ``dopant'' particle is present in the HND, the excitation energy is transferred to the dopant, which in turn is ionized.
This process is called Penning ionization (PEI)~\cite{frochtenicht_photoionization_1996} and can be regarded as a member of the family of interatomic Coulombic decay processes.~\cite{ben_ltaief_charge_2019,Jahnke:2020}
When the photon energy exceeds the ionization energy of helium ($h\nu >24.6$~eV), a positive hole (He$^+$ ion) is created in the HND.
This positive hole migrates by resonant charge hopping and eventually localizes by binding to a neutral helium atom or by ionizing the dopant by charge transfer ionization (CTI).~\cite{peterka_photoionization_2006, buchta_charge_2013, mandal_penning_2020} 

In this paper, we present a detailed investigation of the fragmentation of H$_2$O, D$_2$O, and O$_2$ clusters embedded in HNDs at two characteristic photon energies below and above the ionization energy of helium (24.6~eV).
At $h\nu = 21.6$~eV, HNDs are resonantly excited and the embedded dopant clusters are ionized by PEI; at $h\nu = 26$~eV, HNDs are directly photoionized, and the dopants are ionized by CTI to the ionized HND host (see SI~Fig.~S9 for an estimate of the contribution of direct ionization of water clusters in HNDs).
Owing to the narrow energy bandwidth of the radiation, these two ionization channels are selectively addressed as opposed to other experiments using electron bombardment, where electrons with kinetic energy $>23~$eV always both impact ionize and excite the HNDs to a certain extent given by the respective cross sections.~\cite{buchta_charge_2013, mandal_penning_2020, ralchenko2008electron}
Thus, in the present experiment, we are able to quantitatively compare the CTI and PEI processes in terms of their efficiencies and the degree of fragmentation of the produced dopant-cluster ions.
Compared to CTI, a significant softening effect is observed for PEI. The observed mean fragment cluster size from PEI is almost doubled compared to CTI, indicating less fragmentation in PEI.
Thus, PEI of doped HNDs could be used as a method to produce weakly-bound, metastable ionic molecular complexes and clusters for spectroscopy and the study of ion-molecule reactions in the gas phase.

\begin{figure*}
\includegraphics[width= 16 cm]
{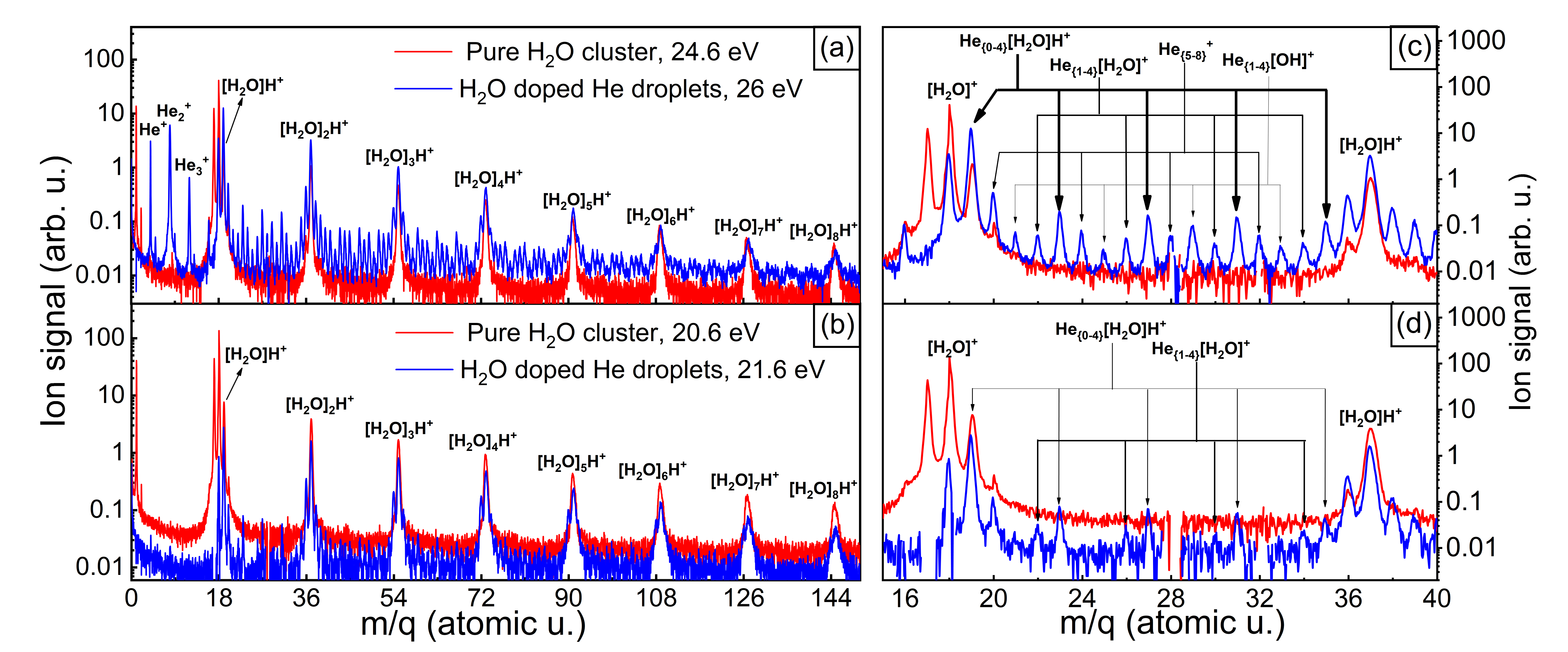}
\caption{\label{fig: Low_dopingwide} Ion mass
spectra of pure H$_2$O clusters and of H$_2$O
clusters formed in HNDs for $\langle N\rangle = 14700 $ and a doping pressure $p_\mathrm{H_2O} = 5\times 10^{-5}$~mbar recorded by CTI following HND photoionization at $h\nu = 26$~eV [blue lines in (a) and (c)]
and by PEI after HND resonant excitation at $h\nu = 21.6$~eV [blue lines in (b) and (d)].
Red lines represent ion mass spectra of pure H$_2$O clusters photoionized at $h\nu = 24.6$~eV [(a) and (c)] and at $h\nu = 20.6$~eV [(b) and (d)]. Panels (c) and (d) are close-ups in the range $m/q = 15$ to 40, showing the structure of protonated and ``unprotonated'' H$_2$O cluster ions.}
\end{figure*}
\section{Experimental methods}
The experiment was conducted at the AMOLine of the ASTRID2 synchrotron, Aarhus University, Denmark.~\cite{pedersen_electronion_2023} The details of the experimental endstation are given in Ref.~\cite{bastian_new_2022} Briefly, the experimental endstation can be divided into three parts.

In the first section, a source chamber contains a cryogenically cooled helium droplet source in which HNDs are generated by supersonic expansion of highly pressurized ($30$~bar) helium gas (He $6.0$) into a vacuum chamber through a small nozzle with diameter $5~\mu$m. 
In our experiments, we varied the nozzle temperature ($T_0$) from 
$11$ to $18$~K, at which HNDs of 
sizes from $\sim2\times10^4$ to 
$\sim2\times10^3$ helium atoms 
per droplet on average are produced, respectively. 
As droplet condensation proceeds out of the gas phase, the size distribution of HNDs follows a log-normal (``LogN'') distribution. 
A skimmer separates the first and second vacuum chambers and skims the central part of the jet where the concentration of droplets is highest. 
Moreover, pure water clusters are also produced to enable comparison between water clusters doped in HNDs and pure water clusters formed by supersonic expansion of water vapor. 
The water cluster source resembles the one of Ref.~\cite{forstel_source_2015}
For producing pure water clusters, we heated the water reservoir to 95\,$^{\circ}$C and the nozzle (80~$\mu$m diameter) to 130\,$^{\circ}$C without using any carrier gas.

In the second section, the HND beam passes through a doping cell (10~mm length) where dopant atoms or molecules are picked up by the droplets.
We injected H$_2$O, D$_2$O, and O$_2$ gas into the doping cell in a controlled way through a fine dosing valve.
The doping cell pressure, as measured by a gas-type independent pressure gauge, was maintained between $5\times10^{-5}$~mbar and $5\times10^{-4}$~mbar when doping with H$_2$O and at $5\times10^{-4}$~\textrm{mbar} when doping with D$_2$O and O$_2$.
A mechanical chopper placed between the first skimmer and the doping cells periodically interrupts the HND beam to discriminate the background signal due to ionization of the residual gas and effusive gas leaking from the doping chamber into the interaction region.
Thus, the HND-correlated signal can be obtained by subtracting the background signal from the doped droplet signal. 

In the third section, the doped HND beam enters the interaction chamber, where the photon beam intersects the HND beam at right angles.
Here, the water clusters doped in HNDs are ionized indirectly by the photoexcited or photoionized HNDs at (i) $h\nu = 21.6$~eV (PEI), and (ii) at $h\nu = 26$~eV (CTI).
For pure water clusters, we choose the photon energies (i) $h\nu = 20.6$~eV and (ii) $h\nu = 24.6$~eV, which are equivalent to the internal energy of the metastable helium atom and the He$^+$ ion in the case of PEI and CTI, respectively.~\cite{mudrich_ultrafast_2020,ben_ltaief_charge_2019}
To suppress higher-order radiation from the undulator, a tin (Sn) filter was inserted in the photon beam for the measurements at $h\nu = 20.6$ and $21.6$~eV, whereas an aluminium (Al) filter was used for the measurements at $h\nu = 24.6$ and $26$~eV (see SI~Fig.~S8 for the filter transmission in this photon energy regime).
The photon-energy resolution of the beamline is $E/\Delta{E}>10^3$. We measure the electron position on the detector with a hexanode delay-line detector and measure the time-of-flight (TOF) of the ions in coincidence with the electrons.

In this paper, we analyze only the electron-ion coincidence mass spectra. All ion mass spectra are normalized with respect to the background N$_2^+$ ion signal, which is proportional to the data acquisition time and the photon flux for a given photon energy.
Single or multiple Gaussian peaks are fitted to get the ion yield contribution of different water cluster ions.
Some general terminologies are used throughout the paper; ``protonated'' refers to both protonated and deuterated water clusters, while ``unprotonated'' refers to both $(\mathrm{H_2O})_k^+$ and undeuterated $(\mathrm{D_2O})_k^+$, and ``water'' refers to both H$_2$O, and D$_2$O. 

\section{Results and discussion}
We start the presentation and discussion of the results by comparing typical ion mass spectra recorded for water clusters embedded in HNDs and pure water clusters, see in Fig.~\ref{fig: Low_dopingwide}.
The embedded water clusters are ionized indirectly following the photoexcitation of the HNDs at photon energies $h\nu = 21.6$~eV [(b),~(d)], and $h\nu = 26$~eV [(a),~(c)];
pure water clusters are ionized at the corresponding photon energies $h\nu = 20.6$~eV [(b),~(d)], and $h\nu = 24.6$~eV [(a),~(c)], respectively.
For pure and HND-embedded water clusters, the formation of protonated water-cluster cations is a dominant process at all photon energies studied here.
Formation of ``unprotonated'' water clusters can be seen up to $k = 2$ for pure water clusters and up to $k = 8$ for water clusters embedded in HNDs (see SI~Fig.~S1). For HND-embedded water clusters, the mass spectra recorded at the two photon energies are clearly dissimilar;
The series of ion mass peaks observed in the mass spectra after CTI [$h\nu = 26$~eV, the blue line in Fig.~\ref{fig: Low_dopingwide} (a) and (c)] are bare helium ions He$_n^+$ ($m/q=4,~8,~12,\dots$~a.~u.) and water cluster fragments complexed with one or a few helium atoms, He$_n$(OH)$^+$
($m/q=21,~25,~29,\dots$~a.~u.), He$_n$(H$_2$O)$_k^+$ ($m/q =18,~22,~26,\dots$~a.~u.), and He$_n$(H$_2$O)$_k$H$^+$ ($m/q =19,~23,~27,\dots$~a.~u.).
In the regime of PEI [$h\nu = 21.6$~eV, blue line in Fig.~\ref{fig: Low_dopingwide} (b) and (d)], mostly He$_n$(H$_2$O)$_k^+$ and He$_n$(H$_2$O)$_k$H$^+$ masses are seen with a weak contribution from water cluster fragments complexed with one or few helium atoms.
No He$_n^+$ ions are formed by PEI because the photon energy falls below the HND autoionization threshold (23~eV).~\cite{buchta2013extreme} The formation of helium complexes around charged clusters, also called Atkins' ``snowballs'',~\cite{atkins_ions_1959,cole_structure_1977,muller_alkali-helium_2009}
results from the balance between the electrostrictive attraction of the helium toward the ion and short-range repulsive exchange forces. 

Remarkably, we do not observe any OH$^+$ in the mass spectra of embedded water clusters; apparently, OH$^+$ formed by CTI is released from HNDs only after the attachment of helium atoms.
An alternative assignment of mass peaks at $m/q  = 20$-$35$ to water cluster fragments attaching multiple H atoms forming (H$_2$O)$_k$H$_l^+$ appears unlikely due to the high dissociation energy of neutral water into the radicals H and OH of about 5~eV per molecule.~\cite{ruscic2002enthalpy}
The series of mass peaks with $m/q=4n+1$~a.~u. is likely due to He$_n$H$^+$ complexes (see also Fig.~S2 in the SI), which is in agreement with an earlier interpretation.~\cite{hogness_ionization_1925, katada2013mass, shibata1973ionizing}
These ions are created by fragmentation of H$_2$O$^+$ into H$^+$ and OH via the lowest energy fragmentation channel.
As the formation of He$_n$H$^+$ is exothermic by $>1.8~$eV, these complexes are efficiently ejected out of the droplets.~\cite{lewis_fragmentation_2005, grandinetti_helium_2004, smolarek2010ir}
This series of peaks is only present in the CTI mass spectra; for PEI we do not observe any He$_n$H$^+$ peak series, indicating less fragmentation. 

When water clusters embedded in HNDs are indirectly ionized by PEI or CTI, the excess energy deposited in the water cluster cations is partly released to the HNDs, leading to the evaporation of helium atoms and rapid cooling of the dopant clusters.
In this way, the fragmentation of water clusters is mitigated, creating ``unprotonated'' water cluster ions.
In contrast, in the case of direct ionization of pure water clusters,  fragmentation is the prevailing processes, yielding mostly protonated water cluster ions.
Thus, indirect ionization of dopants in HNDs can be considered a ``softer ionization'' mechanism than direct photoionization of pure H$_2$O clusters. 

\begin{figure}
\includegraphics[width=11 cm]{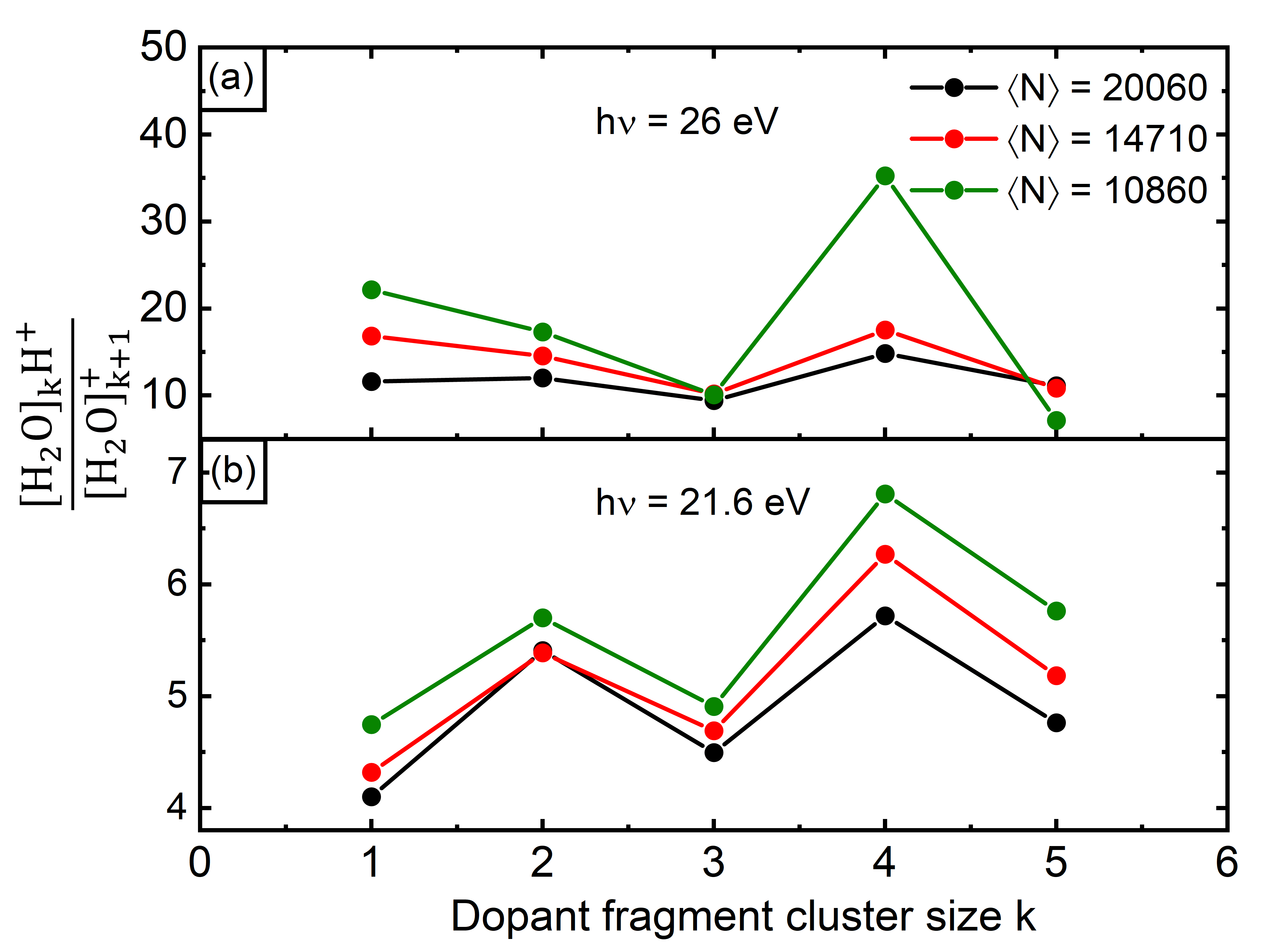}
\caption{\label{fig:Ratio_protonated_to_unprotonated} Comparison of the ratio between yields of protonated water clusters (H$_2$O)$_k$H$^+$ to the corresponding yields of ``unprotonated'' cluster
(H$_2$O)$_{k+1}^+$ ions formed in HNDs of different sizes and at $p_\mathrm{H_2O} = 5\times10^{-5}$~mbar.
Photon energies are $h\nu = 26$~eV in (a) and $h\nu = 21.6$~eV in (b).}
\end{figure} 
The relative degree of fragmentation of dopant clusters induced by CTI and PEI can be expressed quantitatively by the ratio between the protonated and ``unprotonated'' cluster ions.
A higher value of this ratio indicates a higher rate of fragmentation of water clusters induced by CTI, respectively a lower fragmentation rate by PEI.
Fig.~\ref{fig:Ratio_protonated_to_unprotonated} shows the variation of this ratio 
as a function of fragment cluster size $k$ for different HND sizes $\langle N\rangle$.
The fact that this ratio is on average about four times higher in the CTI regime ($h\nu = 26$~eV) compared to the PEI regime ($h\nu = 21.6$~eV) underscores the ``soft ionization'' character of PEI with respect to CTI.
The general trend of a decreasing ratio with increasing HND size (green to black symbols in Fig.~\ref{fig:Ratio_protonated_to_unprotonated}) confirms the concept that fragmentation is suppressed by coupling of the ionized water clusters to the ultracold HNDs. 
The ratio is highest for the known magic number $k = 4$, indicating a high stability of the (H$_2$O)$_4$H$^+$ ion,
which is a building block for larger protonated water cluster ions; a central H$_3$O$^+$  ``eigen cation'' is caged by three H$_2$O molecules forming H$_3$O$^+$(H$_2$O)$_3$.~\cite{adoui_ionization_2009,tsuchiya1989clusters} 
As a measure for the ``soft ionization'' effect of PEI, we define a ``softening coefficient'', $R_k$, in a similar way as in Ref.~\cite{liu_coreshell_2011}
\begin{equation}
  R_k = \frac{\left[\left(\mathrm{H}_2 \mathrm{O}\right)_{k+1}^{+} /\left(\mathrm{H}_2 \mathrm{O}\right)_{k} \mathrm{H}^{+}\right]_{21.6 \hspace{0.05cm}\textrm{eV}}}
{\left[\left(\mathrm{H}_2 \mathrm{O}\right)_{k+1}^{+} /\left(\mathrm{H}_2 \mathrm{O}\right)_{k} \mathrm{H}^{+}\right]_{26 \hspace{0.05cm} \textrm{eV}}}.
\end{equation}
$R_k > 1$ ($R_k < 1$) indicates reduced (enhanced) fragmentation at $h\nu = 21.6$~eV compared to $26$~eV, respectively, whereas $R_k = 1$ means no softening effect.
The variation of $R_k$ with the dopant fragment cluster size $k$ is presented in Fig.~S5 in SI.
This coefficient takes maximum values up to $5$-$6$ for $k = 4$ at low doping conditions and $\langle N \rangle =10860$.
A similar maximum value was measured when comparing CTI of water clusters in HNDs co-doped with oxygen molecules \textit{vs.} CTI of water clusters without any co-dopants.~\cite{liu_coreshell_2011}
In the present experiment (PEI \textit{vs.} CTI), the mean value is $R_k = 2$-$3$, averaging over all $k$ and $\langle N \rangle$ studied here.
\begin{figure}
\includegraphics[width=11 cm]{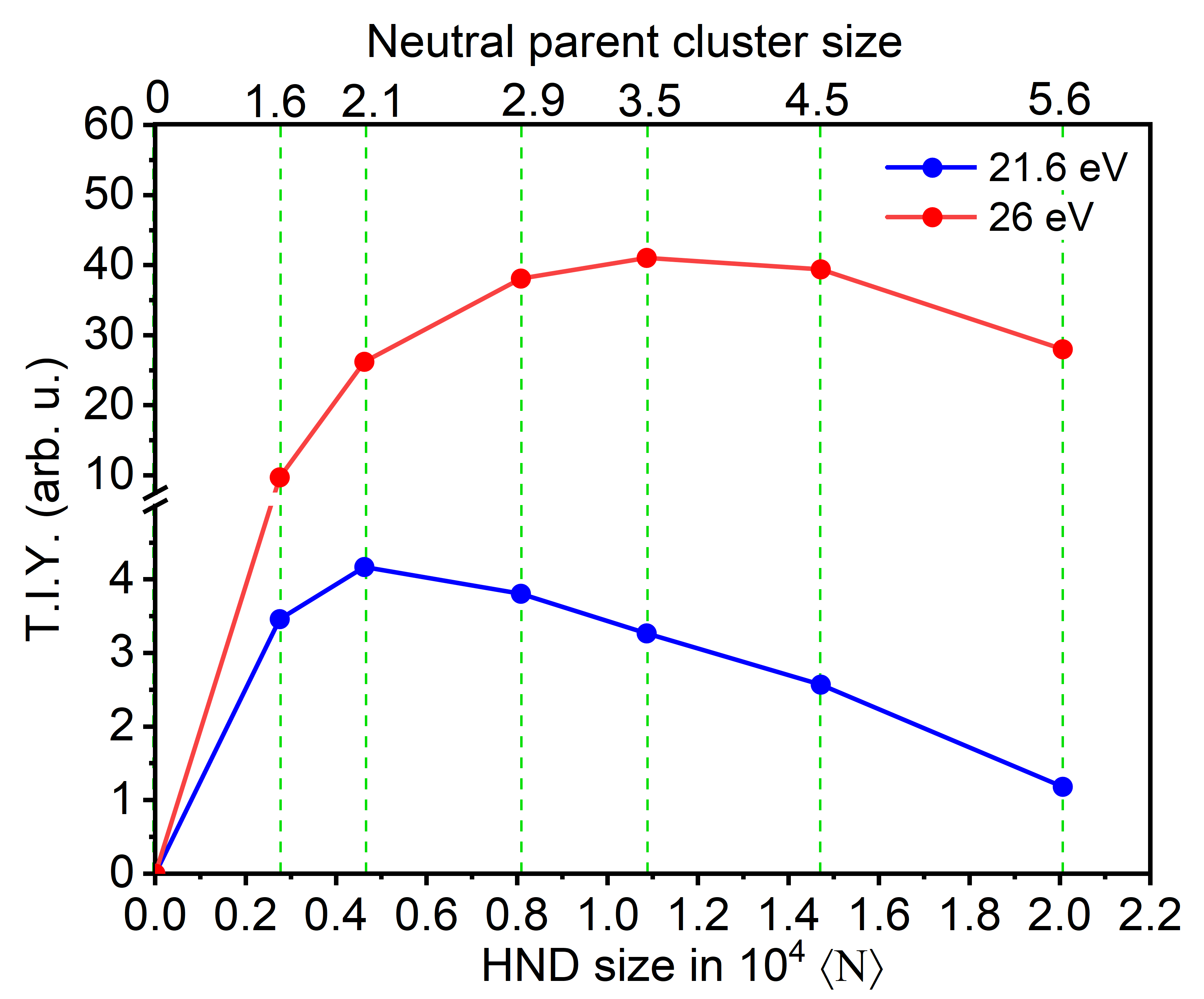}
\caption{\label{fig:Normalized_total_ion_yield} Variation of the total ion yield (TIY) with the HND size for a fixed pressure
$p_\mathrm{H_2O} = 5\times 10^{-5}$~mbar of the water vapour in the doping cell at photon energies $h\nu = 26$~eV (red line) and 21.6~eV (blue line).
The top-axis values indicate the estimated neutral parent cluster sizes from the LogNPoiss distribution; see SI~Fig.~S7 and text for more information.}
\end{figure}
To further quantify the differences between PEI and CTI, we have calculated the total ion yields (``TIY'') by adding up all ion yields of protonated and ``unprotonated'' water cluster ions, including monomer ions, measured by PEI and by CTI.
As we detect ions in coincidence with electrons, the question arises to what extent electrons are emitted from the droplets.
After ionization, some of these could remain attached to anionic droplet fragments. According to our recent study,~\cite{asmussen_dopant_2023} photoelectrons generated at $h\nu = 26$~eV are emitted from HNDs with unity probability up to a droplet size $\langle N \rangle\approx 10^6$ ($R\approx 20$~nm).
As the HNDs used in the present study were smaller than this critical size, it is safe to assume that the TIY defined as the yield of all ions recorded in coincidence with electrons is actually nearly identical to the total yield of emitted ions.
However, the yields of detected free dopant ions produced by CTI and PEI already drop for HNDs with sizes $\langle N\rangle\gtrsim 5\times 10^3$ ($R\approx 5$~nm).~\cite{asmussen_dopant_2023}
The observed reduction of the detected ion yield for larger HNDs was attributed to the limited effective range of 
the He$^+$ photoions migrating through the HND toward the dopant (which is about $2.0$~nm).
For PEI of dopant molecules submerged in the HND interior, an even shorter effective range $\sim 1.1$~nm was found.~\cite{asmussen_dopant_2023}

The TIY of water cluster ions detected as a function of HND size is shown in Fig.~\ref{fig:Normalized_total_ion_yield}.
Overall we note an enhanced CTI efficiency as compared to PEI by about a factor 10. A similar enhancement of CTI over PEI was previously observed for oxygen molecules embedded in HNDs.~\cite{asmussen_dopant_2023}
Due to the shorter effective range of He$^*$ excitations in HNDs and their tendency to be expelled toward the droplet surface and even to detach from the droplets,~\cite{kornilov2011femtosecond,mudrich_ultrafast_2020} dopants submerged in the droplet interior are significantly less efficiently ionized by PEI than by CTI.
In the present case of ionization of dopant clusters in HNDs, the lower TIY measured for PEI might be a further indication that energy transfer to the dopant cluster upon ionization is reduced, thereby lowering the rate of fragmentation;
cold, unfragmented ions tend to stick to the HNDs where they form tightly bound complexes with surrounding helium atoms.

The observed increase of TIY at small HND sizes followed by a decrease toward larger HND sizes can be rationalized by the following partly counteracting trends: \\
(i)	The pick-up efficiency is proportional to the HND cross-section; it increases with HND size leading to more efficient doping.
Therefore both the number of doped HNDs and the dopant cluster sizes increase in the range of small but increasing $\langle N\rangle$.\\
(ii) As the HND size increases further, the efficiencies of dopant ionization by CTI and PEI drop due to the limited effective ranges of He$^+$ and He$^*$ in the HNDs. \\
(iii) Ion ejection probability drops with growing HND size as internally cold ions form tightly-bound complexes with helium inside the HNDs.\\
(iv) In the range of $\langle N\rangle\gtrsim 10^4$, the number density of HNDs in the interaction region decreases, eventually leading to further decreased ion yields.\\
The earlier onset of the drop of TIY for the case of PEI already for $\langle N\rangle>4\times 10^3$ likely reflects the shorter effective range of He$^*$-dopant PEI. Additionally, the colder water-cluster ions formed by PEI are more efficiently trapped in HNDs of growing size.
\begin{figure}[t!]
\includegraphics[width= 12 cm]{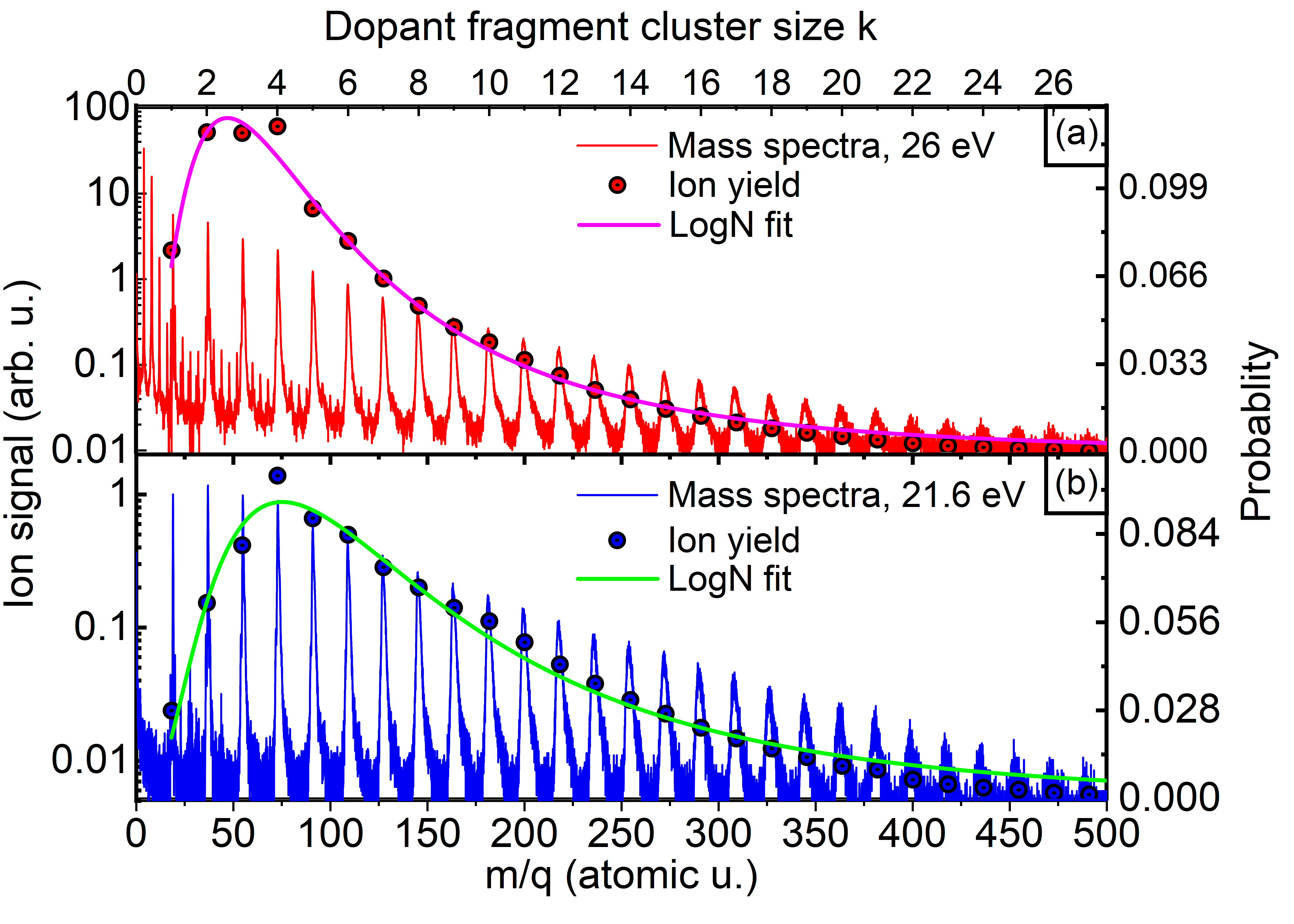}
\caption{\label{fig:High_doping_mass_spectra} Ion mass spectra for H$_2$O clusters doped in HNDs for $\langle N\rangle = 8080$ and $p_\mathrm{H_2O} = 4.5\times10^{-4}$~mbar at photon energies (a) $h\nu = 26$~eV (red lines) and (b) 21.6~eV (blue lines).
Here, the filled circles (red and blue) represent the yields of protonated cluster ions, and the smoothed solid lines (pink and green) represent the best fits of the LogN distributions.}
\end{figure}
A broader overview of the size distributions of fragment water clusters is presented in Fig.~\ref{fig:High_doping_mass_spectra}.
These ion mass spectra were recorded at a higher doping pressure of $4.5\times 10^{-4}$~mbar and at photon energies (a) $h\nu = 26$~eV (red lines) and (b) $21.6$~eV (blue lines).
The nearly complete absence of ``unprotonated'' water-cluster ions in the mass spectra indicates less efficient cooling of the water clusters by the HNDs at these high doping conditions, resulting in the complete loss of OH.
When the heat input due to massive dopant coagulation and fragmentation exceeds the cooling rate provided by the HNDs, fragmentation and ejection of OH occur with high probabilities.

The peak integrals, indicating the abundances of individual cluster sizes, are represented by filled circles (red and blue).
The smoothed solid lines (pink and green) are best fits of a LogN distribution, which nearly perfectly matches the measured abundance distributions.
The clear deviation from the Poissonian distribution, which governs the pick-up statistics assuming a monodisperse and constant HND size,~\cite{lewerenz1995successive,hartmann1996high} is due to (i) the presence of a broad LogN HND size distribution in the experiment;~\cite{harms1998density}
thus, the resulting size distribution of dopant clusters (``LogNPoiss'' distribution: see SI Fig.~S7 and the text for more information) is best described by the product of the dopant size distribution $P_{ck}(N,k)$ (see SI~Fig.~S6 and text for more details) and the HND LogN size distributions in agreement with earlier findings.~\cite{Kollotzek_Efficient_2022}
(ii) The heat of coagulation of dopant molecules leads to a significant shrinkage of the HNDs during the doping process; doping a HND containing $\sim 10^4$ helium atoms with 10 H$_2$O molecules leads to the evaporation of about half of the helium atoms.
The difference between the results from the two fit models, LogNPoiss and LogN distribution, is $<5$\,\%. We conclude that dopant cluster-size distributions are generally well modelled by simple LogN distributions. Therefore, for the rest of this work, we 
used the LogN distribution to interpret our results.
From the LogN fits in Fig.~\ref{fig:High_doping_mass_spectra} we infer the mean fragment cluster sizes $\overline{k} = 5.6$ for CTI at $h\nu = 26$~eV [panel (a)] and $\overline{k}  = 7.3$ for PEI at $h\nu = 21.6$~eV [panel (b)].
The significantly larger fragment-cluster sizes measured by PEI are another manifestation of the ``soft ionization'' character of PEI in HNDs compared to CTI. 

For pure water clusters photoionized at the corresponding photon energies $h\nu=24.6$~eV and $20.6$~eV, the LogN fit yields mean fragment cluster sizes $\overline{k} =2.46$ and $2.45$, respectively (see SI Fig.~S3),
which are identical within the experimental error.
Thus, we can safely exclude the mere energy difference between the two ionization processes in HNDs as the cause of the differing fragment-cluster sizes.
Likewise, an earlier study using electron-impact ionization of pure water clusters found that the fragmentation rate is independent of the electron energy in the range $15$-$90$~eV.~\cite{lengyel_extensive_2014} 
\begin{figure}
\includegraphics[width= 11 cm]{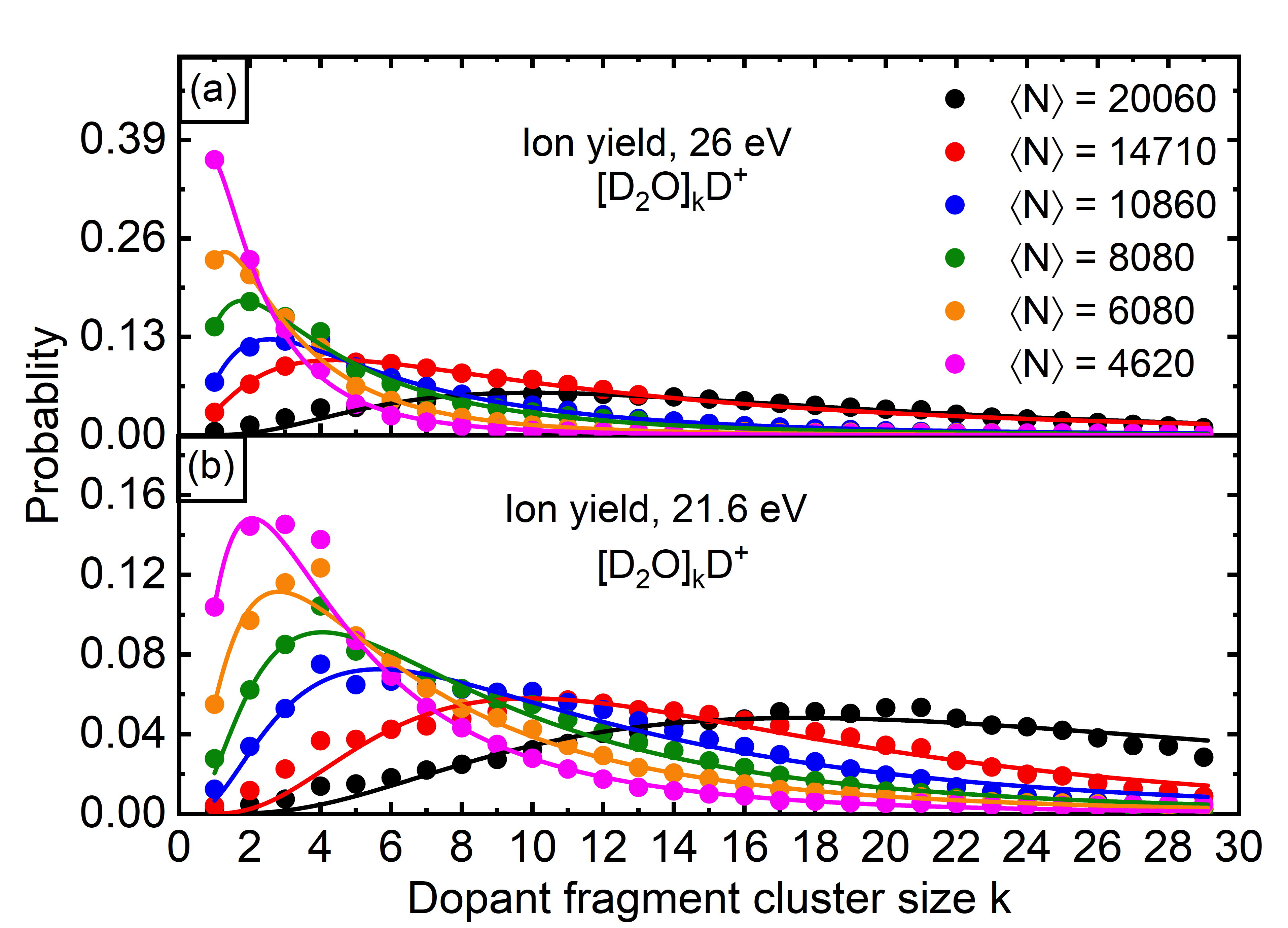}
\caption{\label{fig:D2O_log-normal_dist} Ion yield distributions of D$_2$O clusters formed in HND at $p_\mathrm{H_2O} = 5\times10^{-4}$~mbar for different HND sizes, measured for (a) CTI at $h\nu = 26$~eV, and (b) for PEI at $h\nu = 21.6$~eV.
The filled circles represent the ion yields of the corresponding protonated clusters. The smoothed solid lines represent LogN best fits.}
\end{figure} 
To quantify the soft-ionization effect of the PEI process more systematically, we repeated these measurements for different HND sizes and using D$_2$O as dopants, see Fig.~\ref{fig:D2O_log-normal_dist}.
The choice of D$_2$O instead of H$_2$O has the advantage that protonated and ``unprotonated'' cluster-ion peaks in the mass spectra are better separated. One immediately notices the close connection between the size of the dopant fragment clusters and the size of the HNDs.
Larger HNDs pick up more dopant molecules, which aggregate into larger dopant clusters in the HNDs. We find that the mean fragment cluster size $\overline{k}$ increases nearly linearly with growing HND size $\langle N\rangle$. 
\begin{figure}
\includegraphics[width= 11 cm]{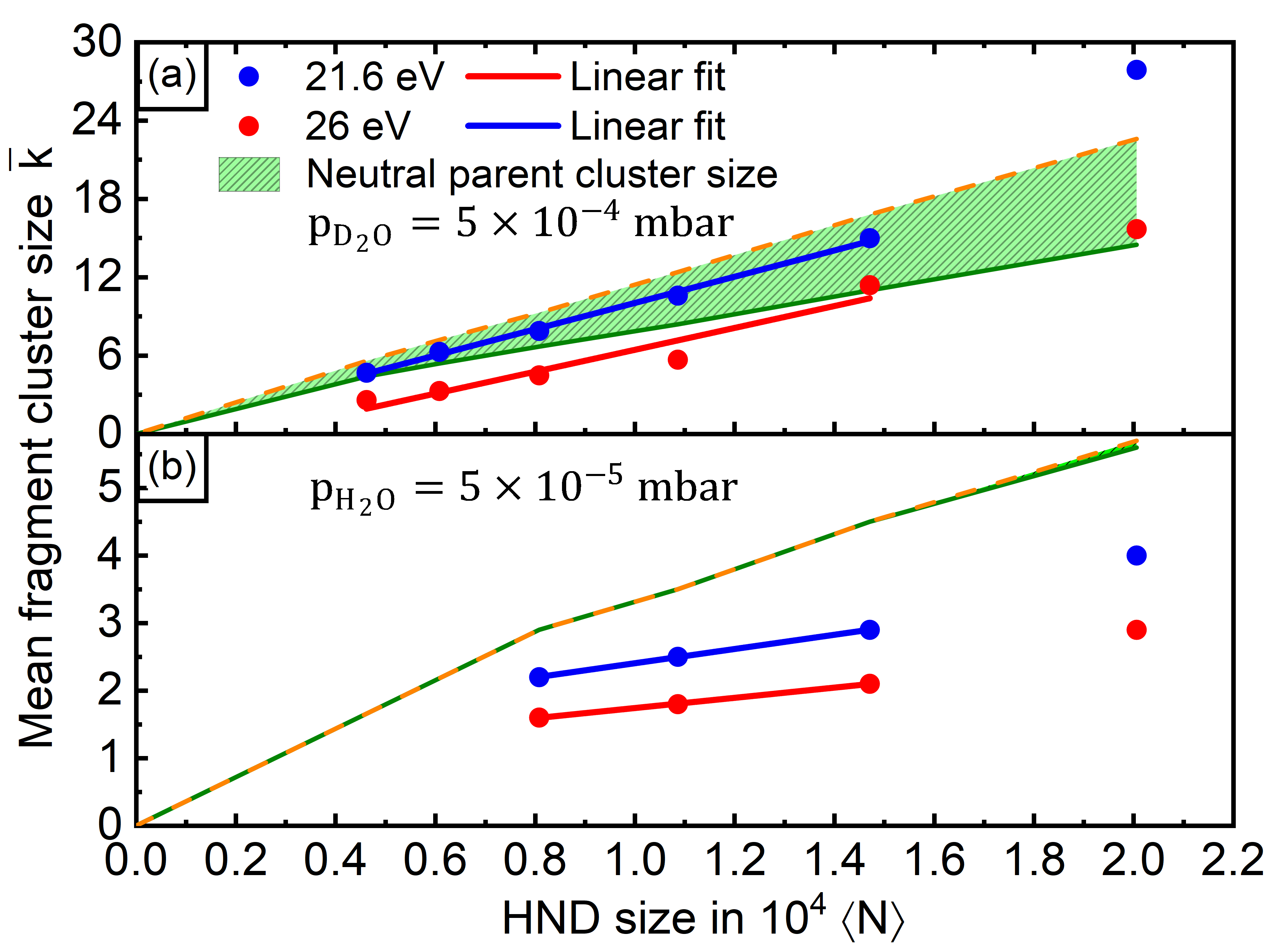}
\caption{\label{fig:Mean_fragment_cluster_size_vs_HND_size} Variation of the mean fragment cluster size $\overline{k}$ of (a) D$_2$O and (b) H$_2$O clusters formed in HNDs at $h\nu=26$~eV (red filled circle) and $h\nu=21.6$~eV (blue filled circle) as a function of the mean HND size.
The solid lines (blue and red) represent the linear fit of the mean fragment cluster sizes.
The shaded green region shows the area covered by the calculated neutral parent cluster sizes from the LogNPoiss distribution in two limiting cases by considering (i) weak dopant-dopant interaction:
binding energy for water dimer (orange dashed line) and (ii) strong dopant-dopant interaction: cluster size dependent binding energies for water cluster (green solid line), respectively.}
\end{figure}
The systematic variation of the mean water fragment cluster size $\overline{k}$ inferred from the LogN fits of the measured size distributions as a function of HND size $\langle N\rangle$ is shown in
Fig.~\ref{fig:Mean_fragment_cluster_size_vs_HND_size} (a) for D$_2$O clusters at high doping conditions, $p_\mathrm{D_2O} = 5\times10^{-4}$~mbar, and (b) for H$_2$O clusters at low doping conditions, $p_\mathrm{H_2O} = 5\times10^{-5}$~mbar.
The approximately linear dependence of $\overline{k}$ on $\langle N\rangle$ mainly reflects the increasing efficiency of larger HNDs to pick up dopant molecules. For D$_2$O doping, the softer PEI leads to nearly twice as large water clusters compared to CTI in the full range of $\langle N\rangle$. Neutral parent cluster sizes are calculated for two limiting cases:
(i) Strong dopant-dopant interaction and (ii) weak dopant-dopant interaction.
The strong dopant-dopant interaction is given when the full equilibrium binding energies are taken into account for each cluster size [see S.~I., equation (3)].
The result is shown as a solid green line in Fig.~\ref{fig:Mean_fragment_cluster_size_vs_HND_size}.
In the case of weak dopant-dopant interaction only the water dimer binding energy [see S.~I., equation (4)] is taken into account.
The result is shown as a dashed orange line in Fig.~\ref{fig:Mean_fragment_cluster_size_vs_HND_size}.
Thus, the neutral parent cluster sizes should lie in the shaded green region provided fragmentation is weak.

The experimentally determined mean cluster sizes measured by PEI for D$_2$O doped HNDs indeed lie in the green shaded region, which
we take as an indication that the water clusters are formed in some weakly bound configuration akin to the cyclic structure discussed by Nauta \textit{et al.}~\cite{nauta2000formation} Of course, fragmentation skews the measured distribution towards smaller $\overline{k}$.
The fact that the measured $\overline{k}$ for the ionized clusters is larger than the one we would expect for the case that the most stable, fully relaxed water clusters form, in spite of fragmentation, is a clear indication that the water clusters aggregate into some weakly bound configurations.
The mean fragment cluster size for $\langle N\rangle = 20060$ deviates somewhat from a linear fit.
The corresponding stagnation temperature lies close to the transition from the subcritical to supercritical regimes of HND expansion, so the produced HND size in that regime is not well defined.
This may be a reason for the deviation of this point from the linear trend.

\begin{figure}
\includegraphics[width= 11 cm]{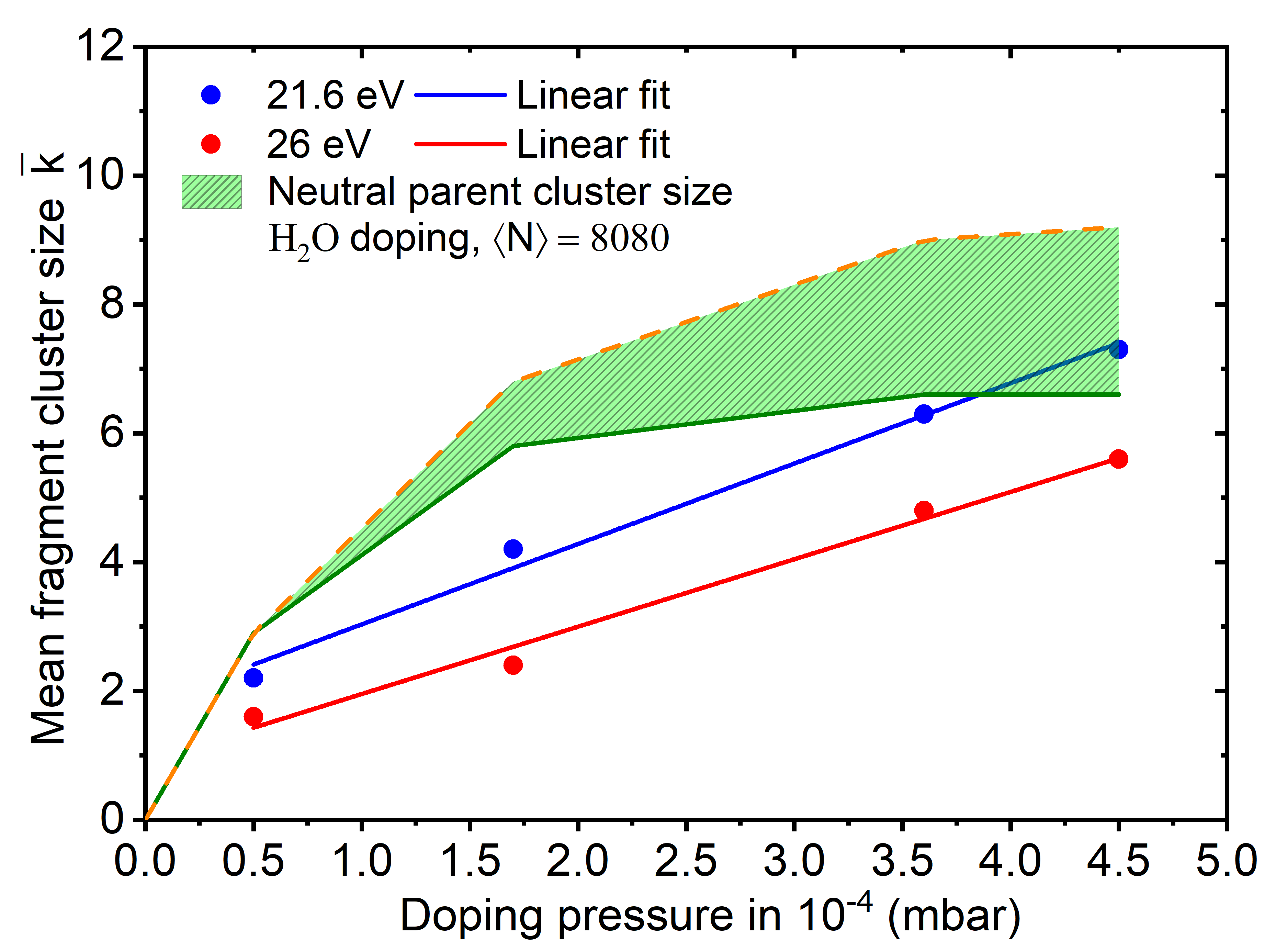}
\caption{\label{fig:mean_fragment_size_vs_doping} Variation of mean water fragment-cluster size as a function of the doping pressure $p_\mathrm{H_2O}$ for $\langle N\rangle = 8080$ at $h\nu = 26$~eV
(red filled circles) and $h\nu = 21.6$~eV (blue filled circles). The solid lines (red and blue) represent the linear fit of the mean fragment cluster sizes.
The shaded green region shows the area covered by the calculated neutral parent cluster sizes from the LogNPoiss distribution in two limiting cases:
(i) Weak dopant-dopant interaction given by the binding energy of the water dimer (dashed orange line), and
(ii) strong dopant-dopant interaction given by cluster size-dependent binding energies for water clusters (solid green line).}
\end{figure}

Similarly to the HND size-dependence, the dependence of $\overline{k}$ on the doping pressure is approximately linear, see Fig.~\ref{fig:mean_fragment_size_vs_doping}. The values of $\overline{k}$ measured by PEI exceed those measured by CTI.
The neutral parent cluster size has been calculated for $\langle N \rangle = 8080$ in a similar way as mentioned before. The calculated neutral parent cluster sizes saturate around $\overline{k} = 7~\mathrm{and}~9$ for strong and weak dopant-dopant interaction, respectively.
This represents the maximum size of the water cluster that can be formed at in HNDs of the size $\langle N \rangle = 8080$, limited by evaporation of all He atoms.

Several experiments~\cite{echt_evolution_1984, kranabetter_uptake_2018, mackie2023magic} have confirmed the values $k= 4,~10,~ 21$ and $28$ as magic numbers in protonated water clusters.
By comparing the ion yield distribution at $h\nu = 26$~eV [Fig.~S4~(a)] and at $21.6$~eV [Fig.~S4~(b)], one can clearly see more pronounced local maxima in the ion yield at magic numbers in the ion-mass spectra recorded at $h\nu =21.6$~eV.
Apparently, the softer PEI tends to enhance the yield of more stable ionic complexes as compared to CTI.

\begin{figure}
\includegraphics[width= 11 cm]{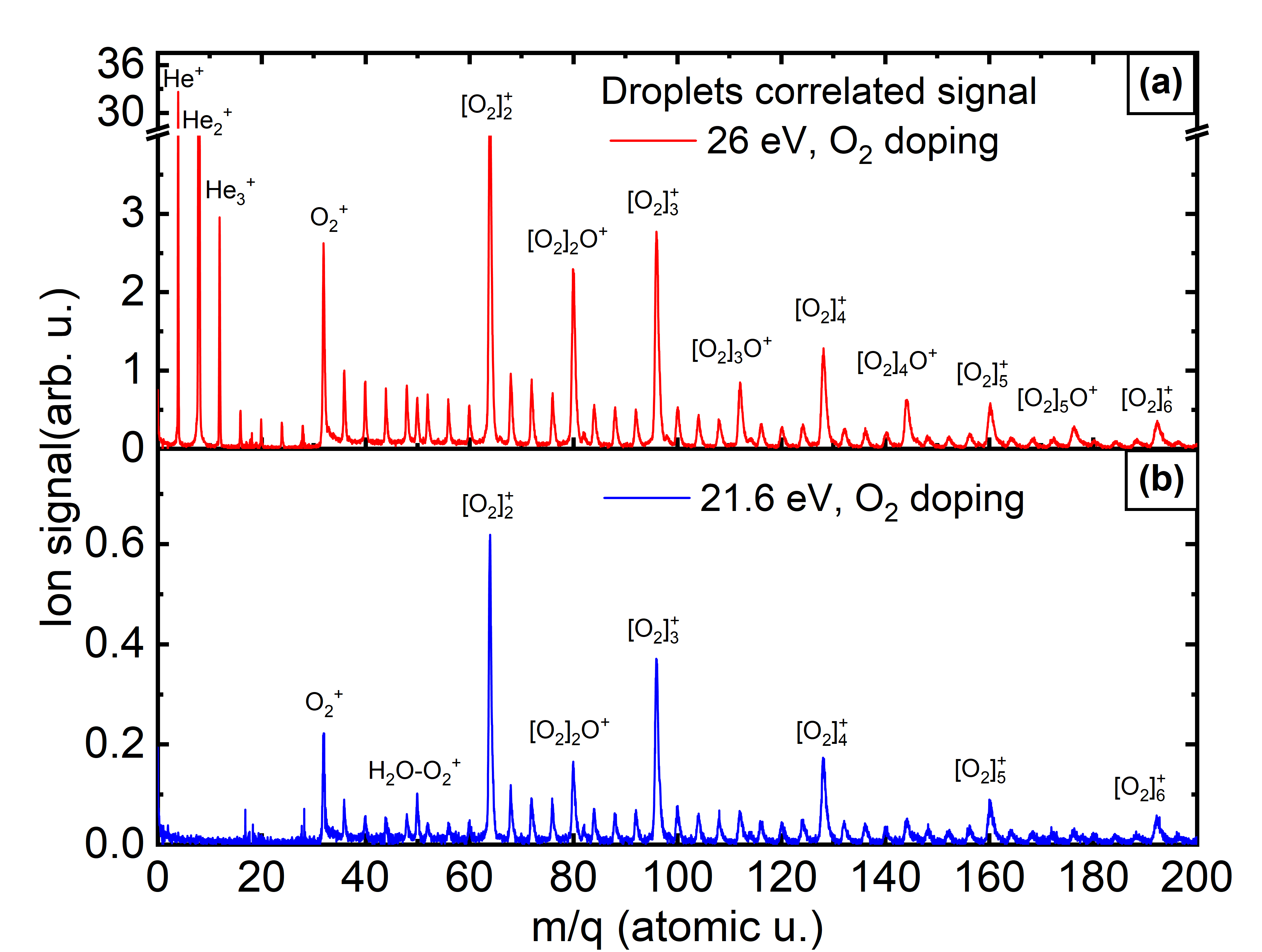}
\caption{
\label{fig:O2_mass_spectra} 
Ion mass spectra of O$_2$ cluster doped in HNDs for $\langle N\rangle = 8080$, and $p_\mathrm{O_2} = 5\times10^{-4}$~mbar at
photon energies  of (a) 
$26$~eV (red line) and
(b) $21.6$~eV (blue line).}
\end{figure}

To assess the general validity of PEI being the softer ionization process, we extended our study to oxygen (O$_2$) clusters formed in HNDs.
Fig.~\ref{fig:O2_mass_spectra} shows mass spectra for O$_2$ doped HNDs with doping pressure $p_\mathrm{O_2}= 5 \times 10^{-4}$~mbar at (a) $h\nu =26$~eV and (b) $21.6$~eV. 
The mass spectra contain both even (O$_{2n}^+$) and odd (O$_{2n+1}^+$) cluster-ion mass peaks with a series of complexes with helium atoms, He$_n$O$_{2n,~2n+1}^+$, with gradually diminishing intensity. 
Generally, CTI again produces higher yields of dopant ions than PEI.~\cite{asmussen_dopant_2023} 
The most striking difference between the two mass spectra is the presence of odd O$_{2n+1}^+$ cluster ions in the case of CTI, which are almost absent in the case of PEI, except O$_5^+$ which is a magic number.~\cite{Shepperson_2011} 
The formation of odd O$_{2n+1}^+$ cluster ions indicates a high energy release upon ionization as the breaking of the bond in O$_2^+$ is endoergic by $5.16$~eV.~\cite{luo2007comprehensive} 
Another possible pathway is the creation of an 
electronically excited O$_2^+$ ion by the ionization 
process, which then interacts with the neighboring O$_2$
molecular units by the intracluster reaction O$_{2n}^+$ + 
O$_2  ~\rightarrow$~O$_{2n+1}^+ +$ O.~\cite{binet_oxygen_1988} 
Comparatively less energy is required for the formation of even O$_{2n}^+$ cluster ions as it only requires breaking of the weak van der Waals bonds between O$_{2}$ units. 
Partial suppression of dissociative ionization in O$_2$ clusters by CTI in HNDs ionized by electron impact has been reported in earlier work.~\cite{Ellis_Helium_2015, Shepperson_2011} 
The nearly complete suppression of O$_2$ bond dissociation in the PEI regime highlights the ``soft ionization'' property of PEI for oxygen clusters as well.
\section{Conclusion}
In this work, we have systematically studied the efficiency of fragmentation induced in molecular clusters by two distinct indirect ionization processes (PEI and CTI in HNDs).
For high doping conditions of the HNDs, protonated water cluster ions are the dominant products for both H$_2$O and D$_2$O clusters, whereas for low doping conditions, ``unprotonated'' water-cluster ions can be
observed up to $k= 8$. Fragment-cluster sizes measured by PEI exceed those measured by CTI by up to a factor of two.
The mean fragment cluster size observed experimentally at PEI for D$_2$O doped HND indicates that water clusters form weakly bound conformation similar to cyclic hexamer structure.
The drastically reduced tendency of inducing fragmentation compared to CTI and direct photoionization of pure water clusters makes PEI a ``soft ionization'' scheme.
This tendency is confirmed for oxygen clusters formed in HNDs; dissociative ionization where the O$_2$ bond is broken is strongly quenched in the PEI scheme as compared to CTI. 

These results are instrumental for mass-spectrometric studies of molecules and complexes and clusters aggregated in HNDs where minimal fragmentation is desired.
The potential of forming metastable, reactive ionic species in the gas phase by PEI of doped HNDs might prove useful for laser spectroscopy of molecular ions and for studies of ion-molecule reactions in the gas phase. The tendency of PEI of causing significantly less fragmentation than CTI can be rationalized as follows:
In the case of CTI, large amounts of energy can be transferred to the products, at least in cases when the potential surfaces of the ionic and the charge transfer states cross,~\cite{stumpf2013efficient} which is the more likely scenario for complex molecules and clusters with many internal degrees of freedom.
In contrast, in PEI a large fraction of the released energy can be carried away by the Penning electron, leaving 
only small amounts of energy deposited in the product ion. This energy redistributes within the ion's various degrees of freedom and may cause its fragmentation. 

In HNDs, both PEI and CTI are affected by the presence of the ultracold helium environment before and after the ionization reaction.
Fragmentation of molecular ions upon CTI has been found to be suppressed due to a cage effect and fast (``explosive'') quenching of fragmentation as well as slow cooling by evaporation of helium atoms.~\cite{Braun_Imaging_2004,Yang_impact_2006,Yang_alcohol_2006,yang_soft_2005,lewis2004electron,lewis_fragmentation_2005}
For PEI, these effects are less well studied, but a stabilizing effect of the interacting helium on the energetics of the PEI reaction has been evidenced.~\cite{ben_ltaief_charge_2019}
A cage effect is less likely to play a role, but the helium atoms involved in the PEI reaction may still extract energy out of the ionized complex in the exit channel.
Kinetic-energy distributions of emitted PEI electrons are generally massively shifted toward low energies for dopants inside HNDs,~\cite{wang2008photoelectron,shcherbinin2018penning,mandal_penning_2020} even for HND sizes where electron-helium scattering is not expected to significantly impact electron energies.~\cite{asmussen_dopant_2023} More extensive experimental and computational studies are needed to elucidate the molecular dynamics underlying these indirect ionization processes in HNDs.
In particular, the role of the superfluid helium environment in quenching certain energetic processes deserves further investigation.
\begin{acknowledgments}
We gratefully acknowledge financial support by the Danish Council for Independent Research (Grant No. 1026-00299B) and the Carlsberg Foundation. S. K. acknowledges
support from the Indo-French Center for Promotion of Academic Research (CEFIPRA), the DST-DAAD bilateral scheme of Dept. of Science and Technology (DST), Govt. of India and German Academic Exchange Service (DAAD),
Germany Science and Engineering Research Board and Technology Development Board, DST, Govt. of India, This project is supported by the Institute of Excellence scheme of the Ministry of Education, Govt. of India, at Indian Institute of Technology (IIT) Madras through the Quantum Center for Diamond and Emergent Materials.
S. D., K. S., S. K. and M. M. acknowledge support from the Scheme for Promotion of Academic Research Cooperation - Ministry of Education, Govt. of India. K. S. is grateful to the IoE travel scheme of IIT Madras for partial support.
A. U. acknowledges support by the Cluster of Excellence “Advanced Imaging of Matter” of the DFG--EXC 2056--project 390715994.
\end{acknowledgments}

\bibliography{Reference} 

\end{document}


\begin{center}
\textbf{ \large
Supplementary Information\\
Fragmentation of Water Clusters Formed in Helium Nanodroplets by Charge Transfer and Penning Ionization
} \\
S. De~\textit{et al.}
\end{center}

\section{\label{sec:level2} Additional mass spectra and ion yield distribution}
Fig.~S$1$ shows ion mass spectra of H$_2$O clusters doped in HNDs at photon energies of $h\nu = 26$~eV (red line) and $21.6$~eV (blue line), respectively.
Both mass spectra are background subtracted to infer the HND-correlated signal.
The mass spectra show ion yield distributions of protonated and ``unprotonated'' H$_2$O cluster ions for both photon energies on a linear scale.
The faster drop of peak intensities for increasing $m/q$ at $h\nu = 26$~eV compared to $21.6$~eV indicates enhanced fragmentation in the CTI regime \textit{vs.} PEI.
Fig.~S$2$ shows the same mass spectra but in the range $m/q = 0$ - $22.5$~a.~u. in log scale to focus on the He$_n$H$^+$ ions mass peaks, where $n = 1$-$3$.
\begin{figure}[h!]
\includegraphics[width= 10 cm]{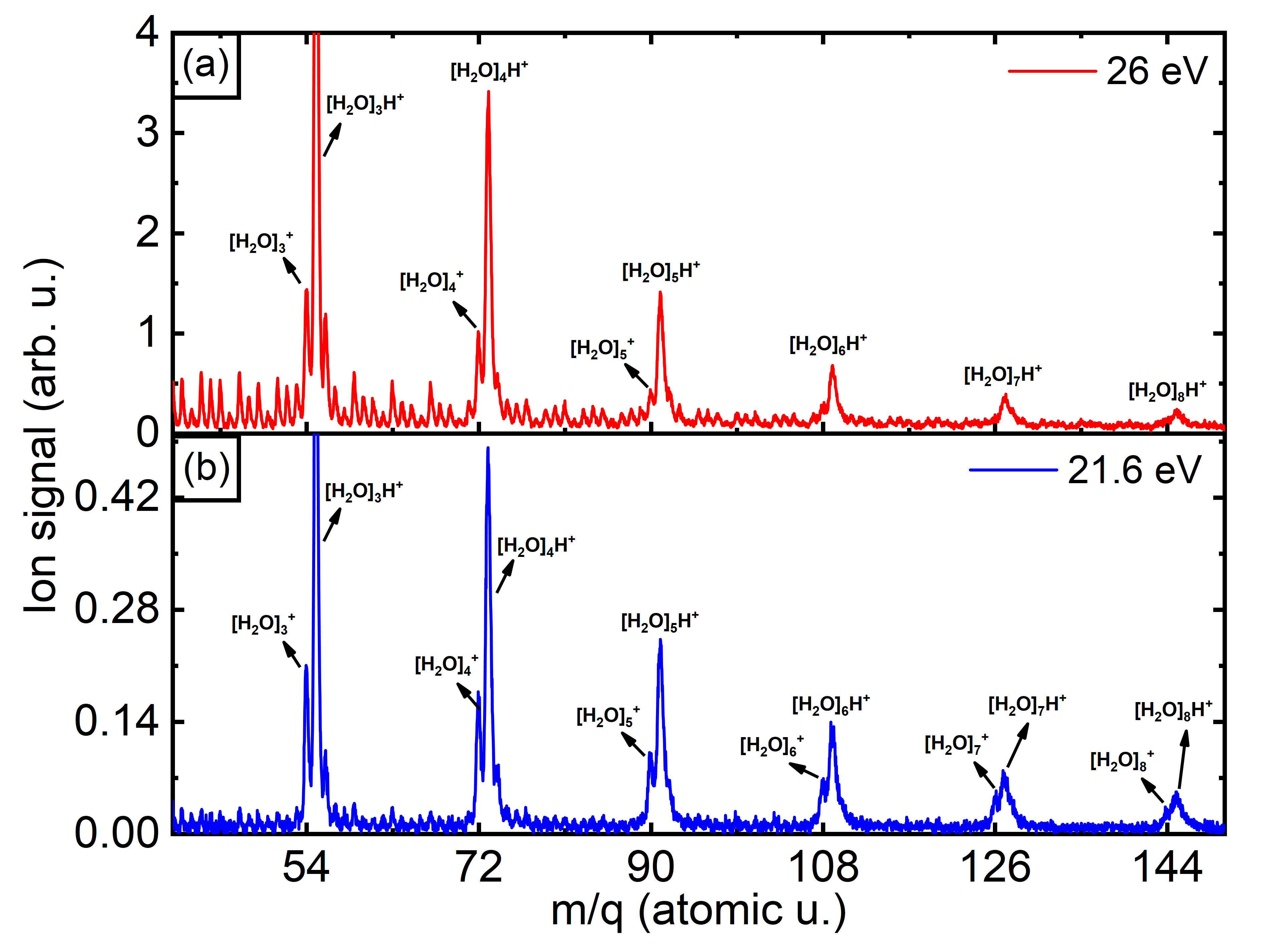}
\caption{\label{{fig: Low_dopingwide}} Ion mass spectra of H$_2$O clusters doped in HNDs at two different photon energies (a) $h\nu = 26$~eV (red lines), and (b) $21.6$~eV (blue lines) for $\langle N \rangle = 8080$ and $p_0 = 30$~bar of He stagnation pressure, showing the distribution of protonated and ``unprotonated'' water cluster ion signals. }
\end{figure}   

\begin{figure}[h!]
\includegraphics[width=0.6\linewidth]{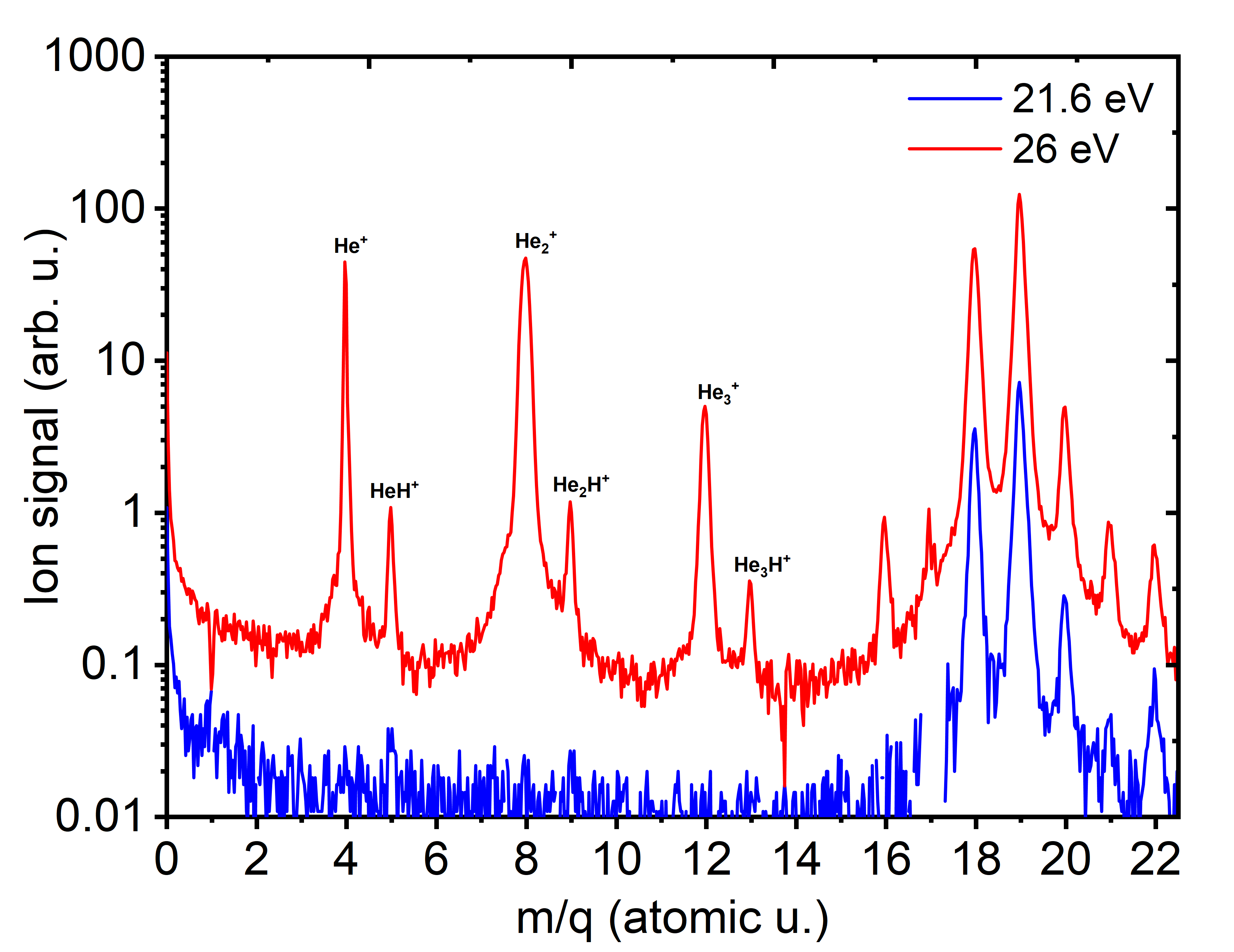}
\caption{\label{{fig: HeH_snowballwide}} Ion mass spectra of H$_2$O clusters doped in HNDs at two different photon energies (i) $h\nu = 21.6$~eV (blue lines), and (ii) $26$ eV (red lines) for $\langle N \rangle = 8080$ and $p_0 = 30$~bar of He stagnation pressure, for $m/q = 0$-$22.5$~a.~u., showing the formation of HeH$^+$, He$_2$H$^+$, and He$_3$H$^+$ ions.}
\end{figure}
\begin{figure}[h!]
\includegraphics[width=0.6\linewidth]{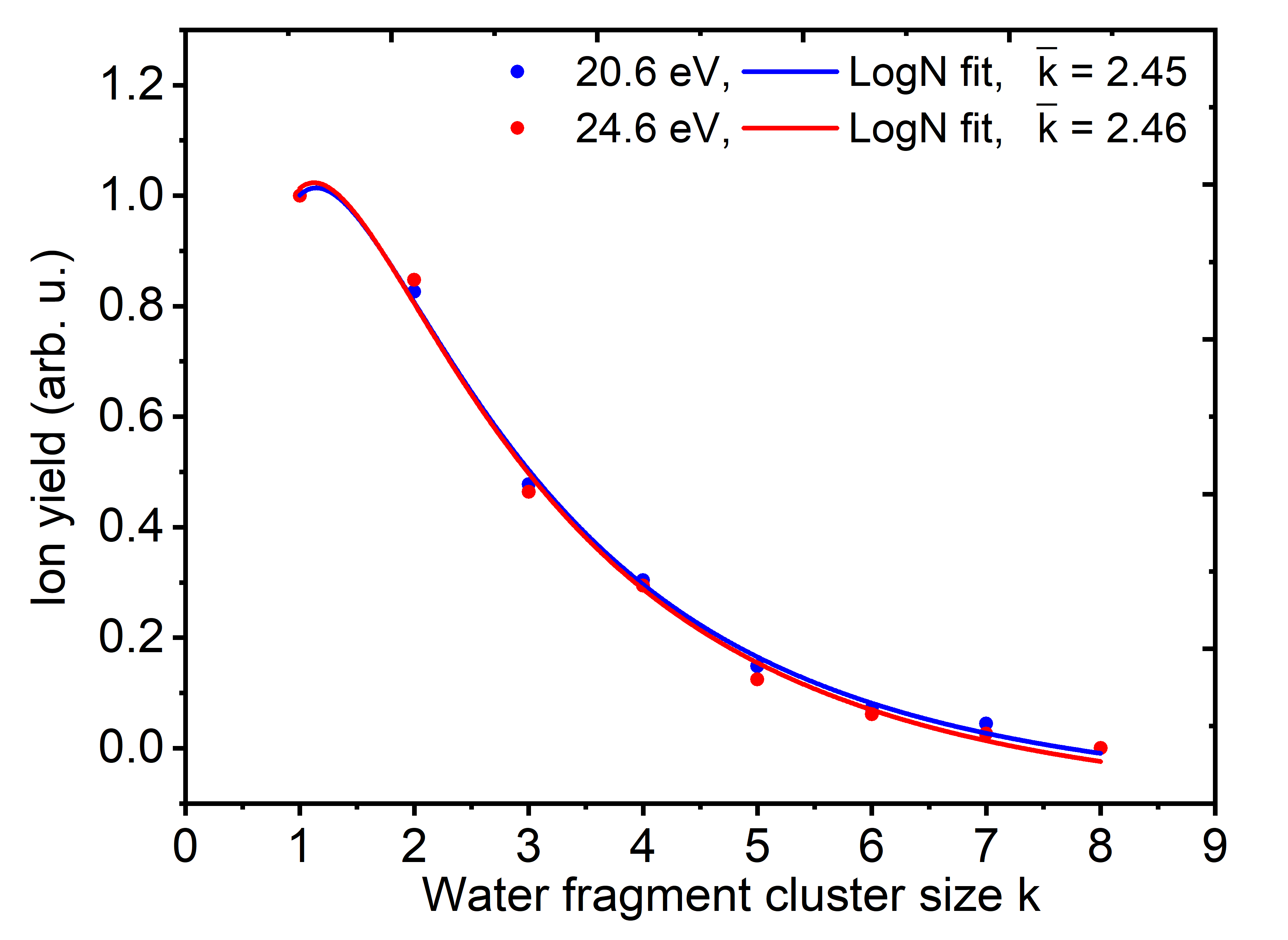}
\caption{\label{{fig: Pure_water_ion_yieldwide}} Ion yield distributions of protonated H$_2$O cluster ion signals for pure water clusters at photon energies of $20.6$ eV (blue circles), and $24.6$ eV (red circles). Solid lines represent the LogN fits of the corresponding ion yield distributions.}
\end{figure} 
In Fig.~S3, the ion yield distribution of the protonated H$_2$O cluster ions signal of pure H$_2$O clusters for photon energies of $h\nu =20.6$~eV (blue circles) and 24.6~eV (red circles) are presented. Here, log-normal (``LogN'') fits of the ion yield distributions show that the mean fragment cluster sizes are almost the same i.~e. the fragmentation of pure H$_2$O clusters is photon-energy independent in this photon energy regime. 

\begin{figure}[hbt!]
\includegraphics[width=0.6\linewidth]{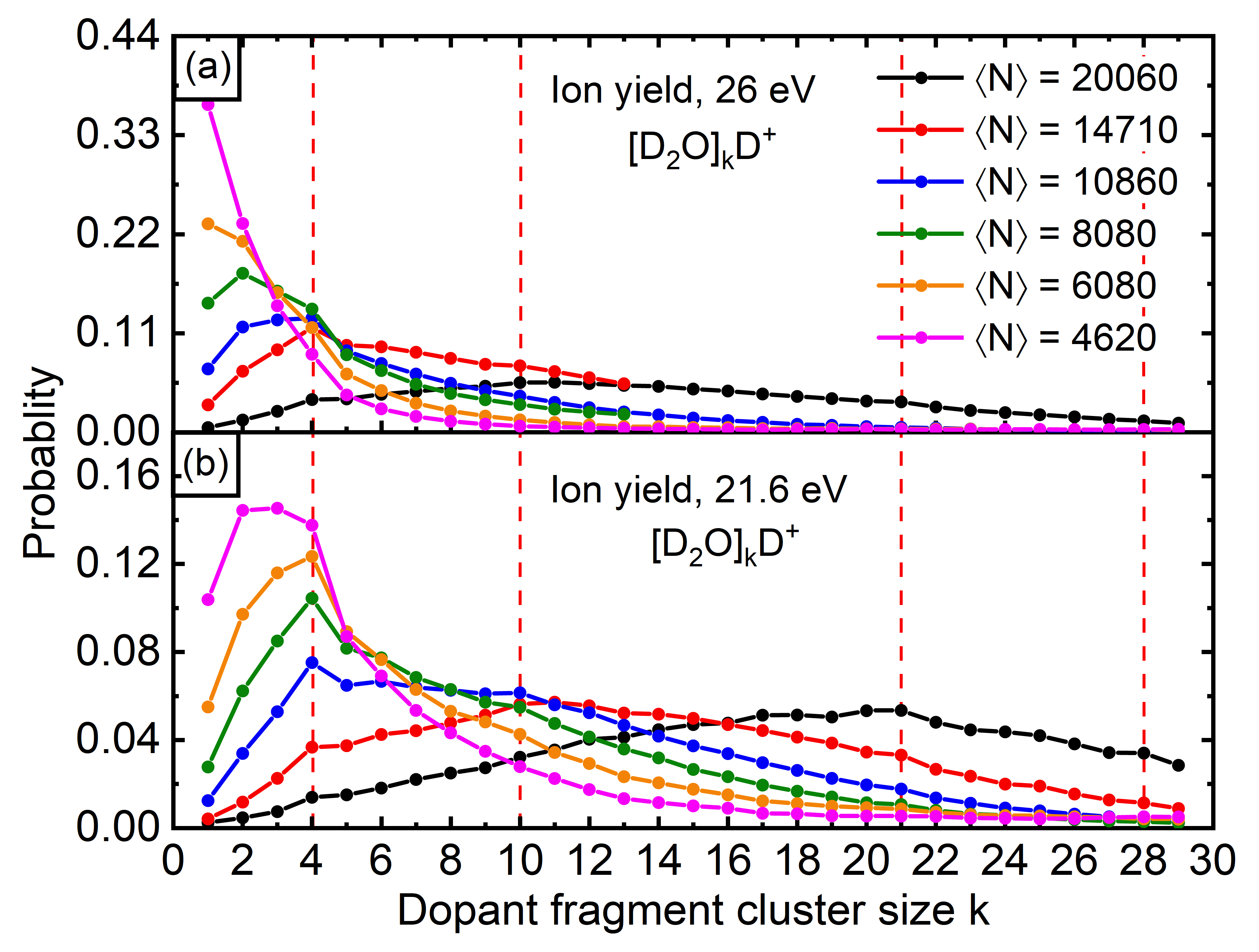}
\caption{\label{{fig: ion}} Ion yield distributions of D$_2$O clusters doped in HNDs at $p_\mathrm{D_2O} = 5\times10^{-4}$~mbar at photon energies (a) $h\nu = 26$~eV and (b) 21.6~eV. Here the filled circles with the solid lines represent the ion yield of the corresponding protonated cluster ions signal. The vertical dashed lines show magic numbers for D$_2$O clusters.}
\end{figure}

By plotting the peak-integrated ion yields as a function of $m/q$ in Fig.~S4, local maxima corresponding to magic numbers are clearly visible.
$\left(\mathrm{D}_2 \mathrm{O}\right)_{4}\mathrm{D}^{+}$, $\left(\mathrm{D}_2 \mathrm{O}\right)_{10}\mathrm{D}^{+}$, $\left(\mathrm{D}_2 \mathrm{O}\right)_{21}\mathrm{D}^{+}$, and $\left(\mathrm{D}_2 \mathrm{O}\right)_{28}\mathrm{D}^{+}$ are the magic numbered water cluster ions, present in protonated ion yield distributions~\cite{Echt_magic_1984,Kranabetter_water_2018, mackie2023magic}. These peaks are more pronounced at $h\nu = 21.6$~eV compared to $26$~eV, showing that the softer
PEI tends to enhance the yield of more stable ionic complexes more efficiently than CTI.

\newpage
\section{\label{sec:level6}Softening coefficient}
The softening effect by HNDs in the fragmentation of water clusters induced by PEI compared to CTI can be expressed quantitatively by the softening coefficient $R_k$.
The variation of $R_k$ as a function of $k$ for different HND sizes is shown in Fig.~S5.
This coefficient has a maximum value of $5-6$ for $k = 4$ at low doping conditions with $\langle N \rangle =10860$ and has an average value of $R_k = 2$-$3$ when averaging over all $k$ and $\langle N \rangle$ studied here. 

\begin{figure}[hbt!]
\includegraphics[width=0.6\linewidth]{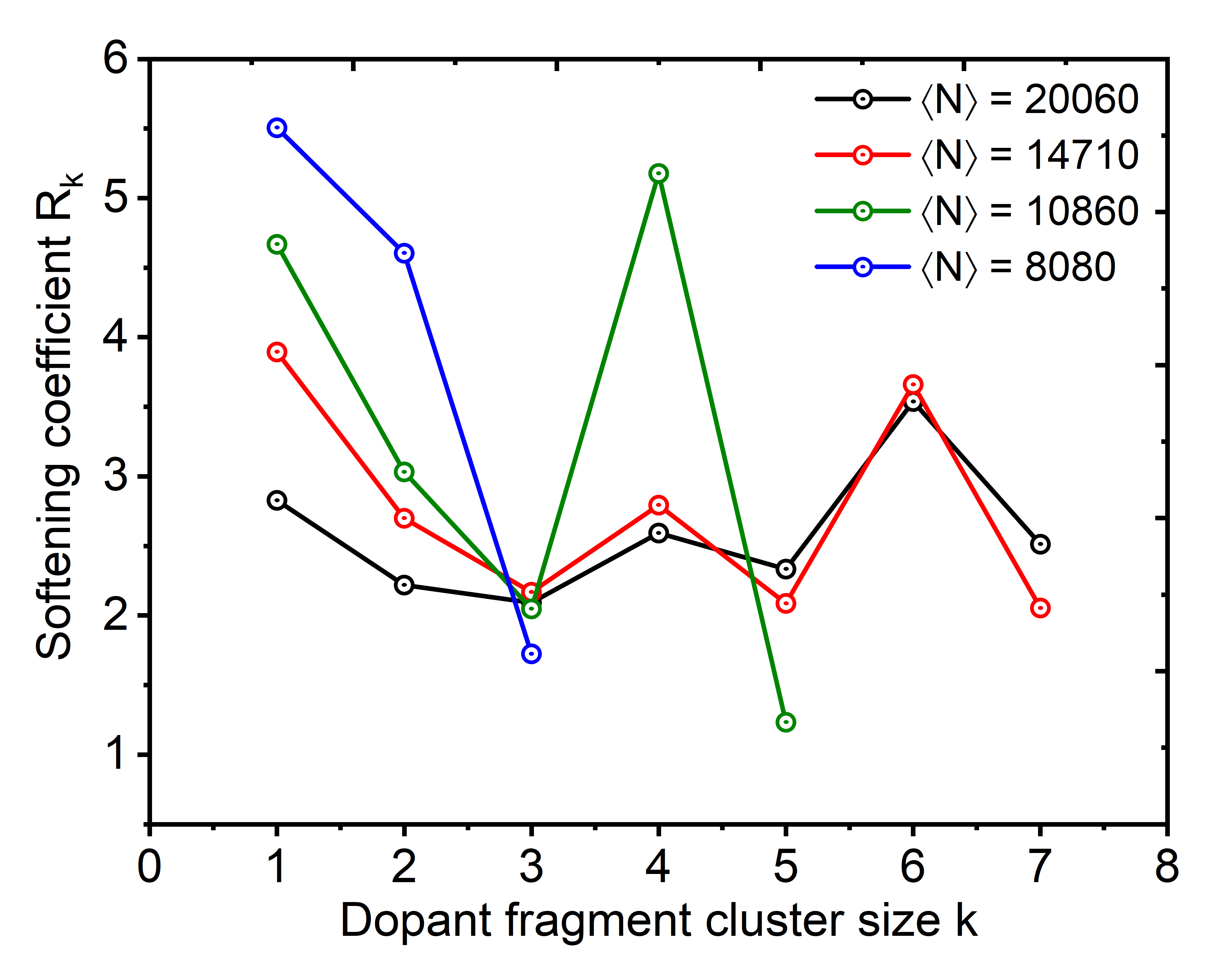}
\caption{\label{{fig: Softenning_coeffwide}} Variation of the softening coefficient with the dopant fragment cluster size for different HND sizes, and $p_\mathrm{H_2O} = 5\times10^{-5}$~mbar for H$_2$O clusters doped in HND. The circles represent the softening coefficient for the corresponding dopant fragment cluster size.}
\end{figure}

\section{Evaporation loss}
Since the binding energy of a He atom $E_b$(He) in a droplet is extremely small (a few K), the energy released during the pick-up of several dopants can evaporate a large number of He atoms from the droplet, resulting in a considerable shrinkage of the droplet size.
The total energy released $(E_{\mathrm{total}})$ due to the pick-up of dopants $X$ can be estimated as
\begin{equation}
   E_{\mathrm{total}} = 
 E_{\mathrm{int}} + E_{\mathrm{coll}} + E_b(\mathrm{He} - X) + E_{b}(X - X) \\
   = 3k_{\mathrm{B}} T + \left(\frac{3}{2} k_{\mathrm{B}} T + \frac{1}{2} m_X v_{\mathrm{cluster}}^2\right) + E_b(\mathrm{He} - X) + E_{b}(X - X).
\end{equation}
Due to its three rotational and three vibrational degrees of freedom, the internal energy of the droplet becomes $E_{\mathrm{int}} =  3k_{\mathrm{B}} T$, where $T$ is 300$~K$.
The collisional energy ($E_{\mathrm{coll}}$) between the dopant and HND can be represented as $\left(\frac{3}{2}k_{\mathrm{B}}T + \frac{1}{2}m_X v_{\mathrm{cluster}}^2\right)$, where
$v_{\mathrm{cluster}}$ represents the velocity of the cluster beam. $v_{\mathrm{cluster}}$ for different HND sizes is retrieved from Ref.~\cite{gomez2011sizes} The used mean HND sizes $\langle N \rangle$ are taken from Ref.~\cite{r5} \\
$E_b(\mathrm{He} - X)$ is the binding energy between He and the dopant molecule~($X$). For the case $X = $H$_2$O, it is $32$~meV~\cite{Lewerenz_1995, green1991calculations}. We assume that $E_b(\mathrm{He} - X)$ value is same for H$_2$O and D$_2$O.
$E_{bk}(X - X)$ is the binding energy liberated for kth dopant pick-up by HND having already doped by (k-1) dopants.
For $X = $H$_2$O, D$_2$O, the cluster size dependent binding energies per molecule are calculated from Ref.~\cite{malloum2019structures} by
\begin{equation} \label{eq:sizedepenent}
E_{\mathrm{k}}(X - X) = ak^b \exp\left(- \frac{k}{c}\right) + d
, \mathrm{~in~kcal/mol}.
\end{equation} 
where $a = 17.1080$, $b = - 0.8361$, $c = 106.3737$, and $d = - 12.0984$.
\begin{equation} \label{eq:ksizedepenent}
E_{\mathrm{bk}}(X - X) = kE_{\mathrm{k}}(X - X) - (k-1)E_{\mathrm{k-1}}(X - X)
, \mathrm{~in~kcal/mol}.
\end{equation} 
Since, $E_{\mathrm{1}}(X - X) = 0$, the binding energy for the water dimer is given by
\begin{equation} 
\label{eq:ksizedepenent}
E_{\mathrm{b2}}(X - X) = 2E_{\mathrm{2}}(X - X)
, \mathrm{~in~kcal/mol}.
\end{equation} 
For the strong dopant-dopant interaction, $E_{b}(X - X) = E_{\mathrm{bk}}(X - X)$, and weak dopant-dopant interaction $E_{b}(X - X) = E_{\mathrm{b2}}(X - X)$.

Hence, the number of evaporated He atoms due to the pick-up of one dopant molecule is given by
\begin{equation}
N_{\mathrm{evap}}(N, k) = \frac{E_{\mathrm{total}}}{E_b(\mathrm{He})}.
\end{equation}
$E_b(\mathrm{He})$ is extracted and extrapolated for different droplet sizes from Ref.\cite{stringari1987systematics}

\section{\label{sec:model}Modelling dopant cluster size distribution in He nanodroplets}
When a HND beam passes through the doping cell, it gets attenuated due to collision with the dopants in the scattering chamber. Because of each collision, the dopant will be captured by the pure HND. Therefore, the intensity of a pure HND of size $N$ will decrease, and the intensity of doped HND will increase. The intensity of this attenuated signal of pure HND of size $N$ will decrease with the distance $z$ in the scattering chamber and can be represented by a set of coupled differential equations.
 
\begin{eqnarray} \label{eq:master}
	\frac{dI_0}{dz} &=& -\lambda_0 I_0 \nonumber \\
	\frac{dI_1}{dz} &=& -\lambda_1 I_1 + \lambda_0 I_0 \nonumber \\
    &\vdots\nonumber \\
	\frac{dI_k}{dz} &=& -\lambda_k I_k + \lambda_{k-1} I_{k-1}
\end{eqnarray}
Here $I_k$ is the intensity of HNDs with exactly $k$ dopants
and $\lambda_k$ is the size-dependent inverse mean free path.
This $\lambda_k$ can be defined as
\begin{equation}
	\lambda_k (N) = F_{a_0} \sigma_{coag}(N_{k}) n_X
\end{equation}
Where $n_X = p_X / k_BT$ is number density of dopants.
For velocity independent hard-sphere potential, $F_{a_0}$ is a velocity averaging correction
factor\cite{Lewerenz_1995}, and it can be approximated by
\begin{equation}
	F_{a_0} \approx \frac{\sqrt{v_{cluster}^2 + v_X^2}}{v_{cluster}}
\end{equation}
$\sigma_{coag}$ is a measure of the probability that the $k$th dopant is picked up by HND containing $k-1$ dopants and undergoes coagulation with the previously trapped $k-1$ dopants. 
\begin{equation}
	\sigma_{coag}(N_{k}) \approx 2 \pi r_0^2 N_k^{2/3}
\end{equation}
Here, we have taken an average cross-section of  $2 \pi r_0^2 N_k^{2/3}$ instead of $ \pi r_0^2 N_k^{2/3}$ or $4 \pi r_0^2 N_k^{2/3}$ due to comparable velocities of HND and dopants. $N_k$ is the size of HND after getting doped by $k$ dopants.
\begin{equation}
	N_k = N_{k-1} - N_{evap} (N_{k-1}, k)
\end{equation}

Hence, the size-dependent inverse mean free path can be represented by
\begin{equation}
	\lambda_k (N_k) = \frac{p_X}{k_B T} \frac{\sqrt{v_{cluster}(N_k)^2 + v_X^2}}{v_{cluster}(N_k)} 2 \pi r_0^2 N_k^{2/3}.
\end{equation}

To solve the set of coupled differential equations (\ref{eq:master}), we can write these as a matrix equation
\begin{equation} \label{eq:masmatrix}
	\frac{d\bold{I}}{dz} = \Lambda \bold{I}
\end{equation}
where, 
\begin{equation}
\Lambda = 
\begin{bmatrix}
    - \lambda_{0} & 0 & \dots & 0 \\
     \lambda_{0} & - \lambda_{1} & \dots & 0 \\
    \vdots & \vdots & \ddots & \vdots \\
    0 & \dots & \lambda_{k-1} & -\lambda_{k} \\
\end{bmatrix}
\end{equation}
Here $\Lambda$ is a $(k+1)~\times~(k+1)$ matrix with $k$ as maximum dopant size.
Now, equation (\ref{eq:masmatrix}) can be numerically solved using the 
fourth-order Runge-Kutta method~\cite{Runge_Kutta_methods}. We assume that at $z = 0$, the intensity has only contribution from pure HND, and the contribution of doped droplets is zero, $I_{k>0} = 0$. We then numerically computed the contributions of doped HNDs at every $\Delta z$ step size ($10~\mu$m). The intensity distribution of doped HNDs at the final step of the Runge-Kutta method is the dopant cluster size distribution $P_{ck}(N,k)$. The deviation of $P_{ck}(N,k)$ from the Poisson distribution, which derives from the shrinkage of the HNDs in the course of the doping process, can be seen in Fig.~S6, where $P_{ck}(N,k)$ (with droplet shrinkage, when $N_{evap} \neq 0$) is represented by solid black line and the Poisson distribution (no droplet shrinkage is considered here) is represented by dashed blue line. Without droplet shrinkage, $P_{ck}(N,k)$ returns to the Poisson distribution as shown by dashed red lines. Due to droplet shrinkage, the peak position is shifted towards smaller dopant cluster size and becomes narrower compared to the Poisson distribution.

\begin{figure}[hbt!]
\includegraphics[width=0.6\linewidth]{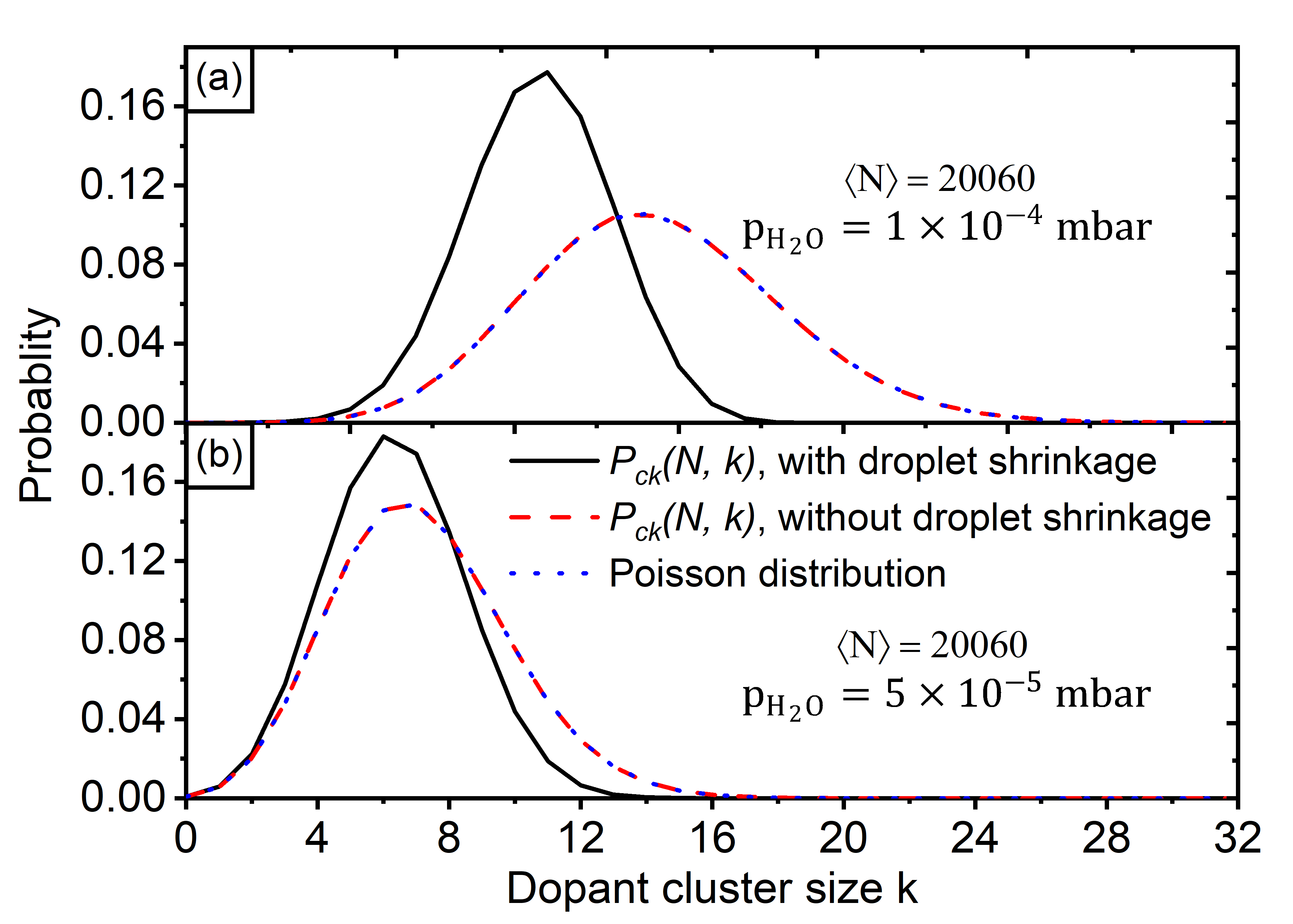}
\caption{\label{{fig: Dopant_cluster_size_distribution_compare}}  The dopant cluster size distribution $P_{ck}(N,k)$ for $\langle N \rangle = 20060$, at doping pressure of (a) $p_\mathrm{H_2O} = 1\times10^{-4}$~mbar (upper panel), and (b) $p_\mathrm{H_2O} = 5\times10^{-5}$~mbar (lower panel) are represented by solid black lines (with droplet shrinkage), and dashed red lines (without droplet shrinkage). The dashed blue lines are the calculated Poisson distributions for the same $\langle N \rangle$ and $p_\mathrm{H_2O}$. $P_{ck}(N,k)$ are calculated by considering cluster size dependent binding energies from equation (\ref{eq:sizedepenent}).}
\end{figure}

\section{Calculation of neutral dopant size}
The size of HNDs produced in subcritical expansion can be described by the log-normal (LogN) distribution function given by 
\begin{equation}
P_\mathrm{LogN} (N, \langle N\rangle, S) = \frac{1}{\sqrt{2\pi} \sigma N}\exp\left[-\left(\frac{\ln N - \mu}{\sqrt{2}\sigma}\right)^2\right].
\end{equation}
As we only know $\langle N \rangle$, we have calculated $S$ such that the difference between full width at half maximum (FWHM) and $\langle N \rangle$ is minimum. To do this, we used the least square fitting routine to minimize the difference.\\
Here, $\sigma$ is the logarithm of the geometric standard deviation 
\begin{equation}
\sigma = \sqrt{\ln\left(1 + \left(\frac{S}{\langle N\rangle}\right)^2\right)}
\end{equation}
and $\mu$ is the logarithm of the geometric mean
\begin{equation}
\mu = \ln(\langle N\rangle) - \frac{\sigma^2}{2}
\end{equation}
The total probability (``LogNPoiss'') distribution is given by the product of dopant size distribution $P_{ck}(N, k)$ and HND LogN size distribution $P_{LogN}(N, \langle N \rangle, S)$,
\begin{equation}
	P_{total}(\langle N \rangle, S, k) = \int_0^\infty P_{LogN}(N, \langle N \rangle, S) P_{ck}(N, k) dN.
\end{equation}
Now, the mean neutral parent cluster sizes are obtained from the median of the LogN fit (median$~(\overline{k}) = e^{\mu'}$, here $\mu'$ is the logarithm of the geometrical mean of LogN fit) of the LogNPoiss distribution. This has been discussed in earlier studies~\cite{bobbert2002fragmentation, lengyel2014extensive}, where they determined mean cluster size from the median of the LogN distribution. The mean fragment cluster sizes $\overline{k}$ are also determined similarly from the median of the LogN fit in the experimental ion yield distribution.

\begin{figure}[hbt!]
\includegraphics[width=0.6\linewidth]{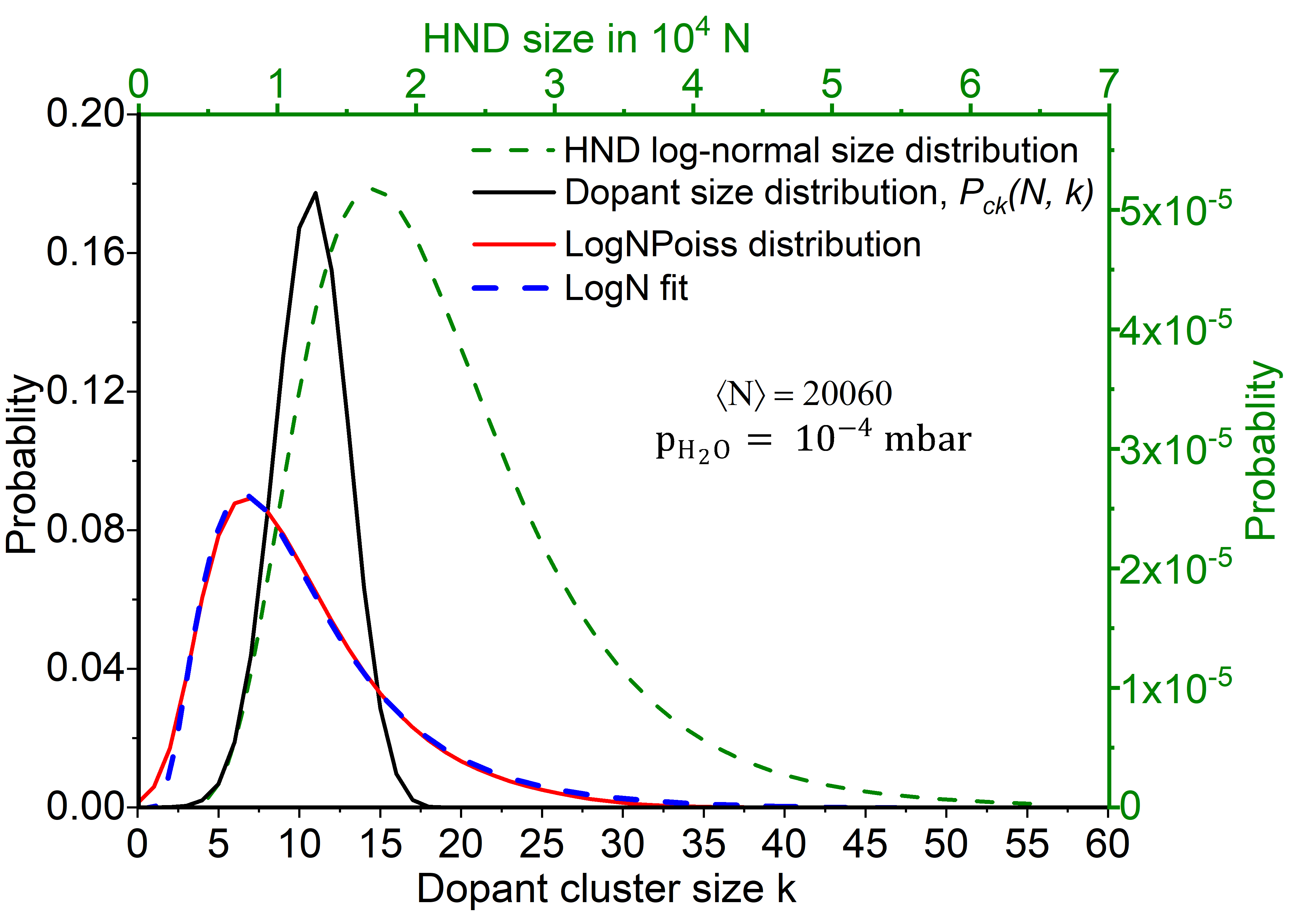}
\caption{\label{{fig: LogNPoiss}}  The LogN size distribution of HNDs for $\langle N \rangle = 20060$, and the dopant cluster size distribution $P_{ck}(N,k)$ at $p_\mathrm{H_2O} = 1\times10^{-4}$~mbar are represented by dashed green line and solid black line, respectively, for H$_2$O cluster doped in HNDs. The total probability (LogNPoiss) distribution is represented by a solid red line. The dashed blue line shows the LogN fit of the LogNPoiss distribution. $P_{ck}(N,k)$ are calculated by considering cluster size dependent binding energies from equation (\ref{eq:sizedepenent}).}
\end{figure} 
Fig.~S7 shows the LogNPoiss distribution (solid red line), HNDs LogN size distribution (dashed green line) and the dopant size distribution for a monodisperse HND size distribution $P_{ck}(N,k)$ (solid black line). The figure also compares the LogNPoiss distribution and the LogN fit (dashed blue line) of the LogNPoiss distribution, showing the similarity between the two models. As the LogNPoiss distribution includes the HND LogN size distribution and the droplet shrinkage effect, the maximum of the distribution will be shifted towards smaller dopant cluster size as compared to the dopant size distribution $P_{ck}(N,k)$. 
\clearpage
\section{\label{sec:level10}Filter transmission}
\begin{figure}[h!]
\includegraphics[width=0.6\linewidth]{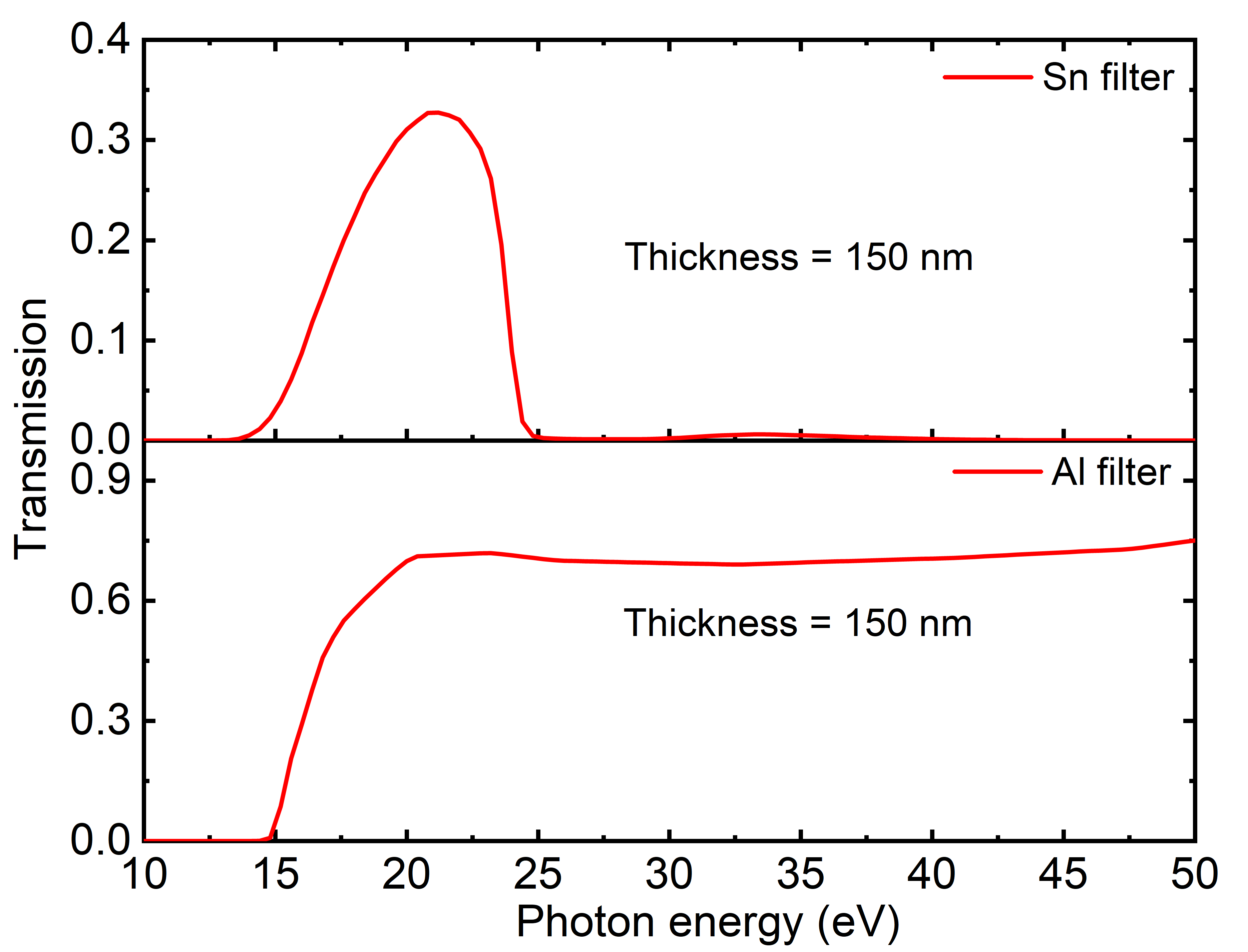}
\caption{\label{{fig: Filter_transmission}}  The calculated filter transmission for Sn filter (Upper panel) and Al filter (Lower panel) for a thickness of $150$~nm at the synchrotron radiation beamline ASTRID2. The calculation is available from an online software~\cite{henke-website}.}
\end{figure}
\section{\label{sec:level10}Direct photoionization of water clusters in HNDs}
\begin{figure}[h!]
\includegraphics[width=0.6\linewidth]
{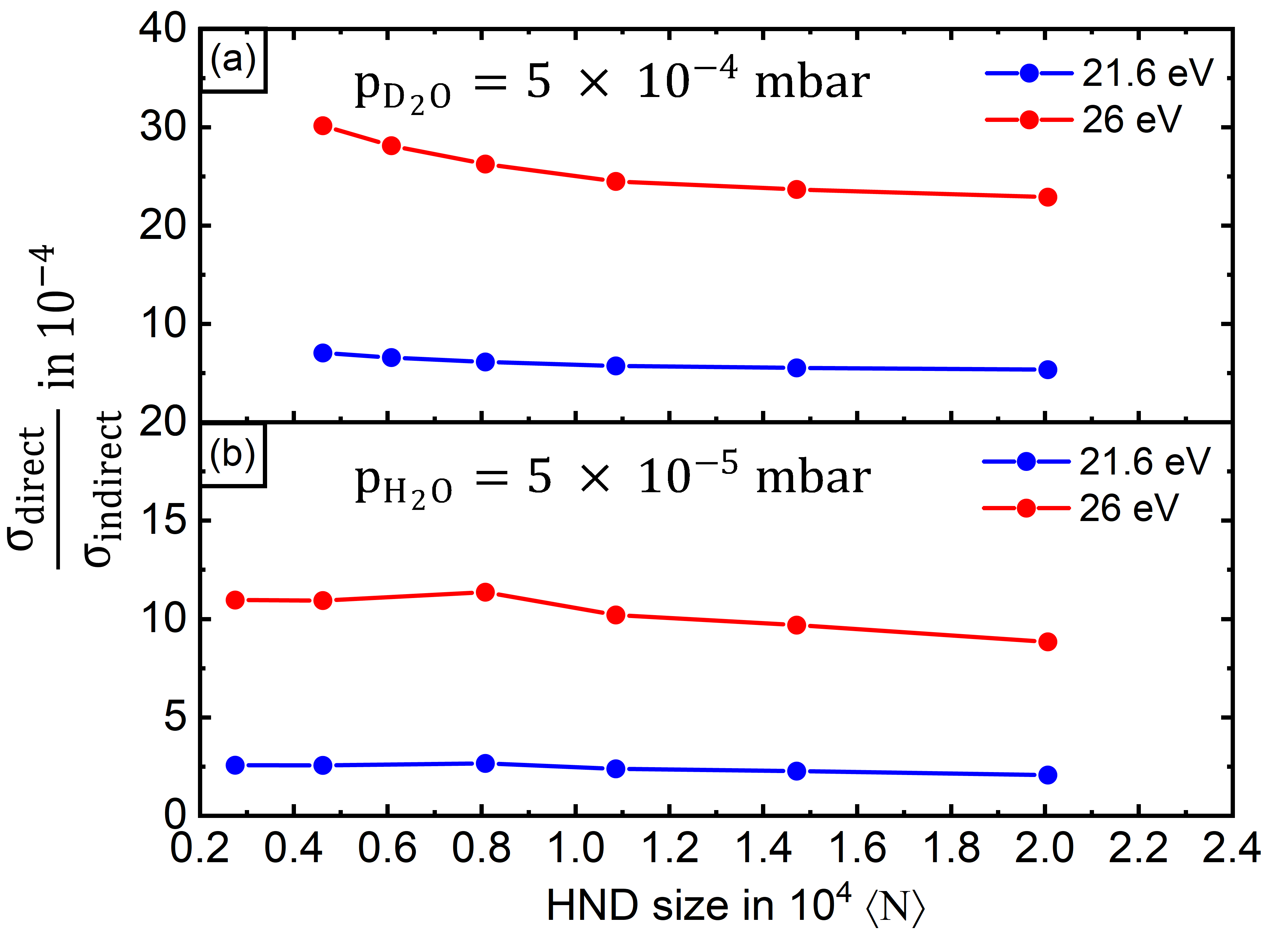}
\caption{\label{{fig: Photoionization_crosssection}}  The variation of the ratio of direct to indirect photoionization cross-section of water clusters doped in HNDs with HND size for (a) D$_2$O (Upper panel), and (b) H$_2$O (Lower panel) respectively. The filled red and blue circles with connecting lines represent the ratio at $26$~eV and $21.6$~eV photon energy, respectively.}
\end{figure}

For direct ionization, the photoionization cross-section per water molecule is $18.5$~Mb at $21.6$~eV and $21.5$~Mb at $26$~eV~\cite{haddad1986total}.\\
For indirect ionization, at 21.6 eV, we have used the photoabsorption cross-section of He atoms, which is $25$~Mb~\cite{buchta2013extreme} and at $26$~eV, we used the photoionization cross-section of He, which is $6.79$~Mb~\cite{samson2002precision}.\\
Hence, the total cross-section for indirect ionization of a HND with a mean size $\langle N \rangle$ is given by
\begin{equation}
\sigma_{\mathrm{indirect}} = \langle N \rangle \times \sigma_{\mathrm{He}}^{21.6~\mathrm{eV},~26~\mathrm{eV}}
\end{equation}
Similarly, the total cross-section for direct ionization of water cluster in HNDs with a mean size $\overline{k}$ is given by 
\begin{equation}
\sigma_{\mathrm{direct}} =\overline{k} \times \sigma_{\mathrm{H}_2\mathrm{O}}^{21.6~\mathrm{eV}, ~26~\mathrm{eV}}
\end{equation}
Fig.~S9 Shows the variation of the ratio of direct-to-indirect ionization cross-section of water clusters doped in HNDs as a function of the HND size. The ratio is comparatively higher at $26$~eV (red lines) for D$_2$O (upper panel) and H$_2$O (lower panel), respectively, than $21.6$~eV (blue lines). We assumed in our calculation that the ionization cross-section for H$_2$O and D$_2$O molecules are the same. As the order of this direct-to-indirect ionization cross-section ratio is very small, the contribution of direct ionization of water clusters in HNDs can be neglected over indirect ionization. 
\newpage
\bibliography{Reference_SI}